\documentclass{aastex631}
\usepackage{amsmath}
\begin{document}
\title{Millimeter Dust Emission and Planetary Dynamics in the HD~106906 System}

\author{Anna J. Fehr}
\affiliation{Astronomy Department and Van Vleck Observatory, Wesleyan University, 96 Foss Hill Drive, Middletown, CT 06459, USA}

\author[0000-0002-4803-6200]{A. Meredith Hughes}
\affiliation{Astronomy Department and Van Vleck Observatory, Wesleyan University, 96 Foss Hill Drive, Middletown, CT 06459, USA}

\author[0000-0001-9677-1296]{Rebekah I. Dawson}
\affiliation{Department of Astronomy \& Astrophysics, Center for Exoplanets and Habitable Worlds, The Pennsylvania State University, University
Park, PA 16802, USA}

\author{Rachel E. Marino}
\affiliation{Astronomy Department and Van Vleck Observatory, Wesleyan University, 96 Foss Hill Drive, Middletown, CT 06459, USA}

\author{Matan Ackelsberg}
\affiliation{Astronomy Department and Van Vleck Observatory, Wesleyan University, 96 Foss Hill Drive, Middletown, CT 06459, USA}

\author{Jamar Kittling}
\affiliation{Astronomy Department and Van Vleck Observatory, Wesleyan University, 96 Foss Hill Drive, Middletown, CT 06459, USA}

\author[0000-0003-2657-1314]{Kevin M. Flaherty}
\affiliation{Department of Astronomy and Department of Physics, Williams College, Williamstown, MA 01267, USA}

\author[0000-0002-7484-5124]{Erika Nesvold}
\affiliation{8336 Dubbs Drive, Severn, MD 21144, USA}

\author[0000-0003-2251-0602]{John Carpenter}
\affiliation{Joint ALMA Observatory, Alonso de C\'{o}rdova 3107, Vitacura, Santiago, 763 0355, Chile}

\author[0000-0003-2253-2270]{Sean M. Andrews}
\affiliation{Harvard-Smithsonian Center for Astrophysics, MS-42, Cambridge, MA 02138, USA}

\author[0000-0003-3017-9577]{Brenda Matthews}
\affiliation{University of Victoria, 3800 Finnerty Road, Victoria, BC, V8P 5C2, Canada}
\affiliation{National Research Council of Canada Herzberg, 5071 West Saanich Road, Victoria, BC V9E 2E7, Canada}

\author[0000-0003-4909-256X]{Katie Crotts}
\affiliation{University of Victoria, 3800 Finnerty Road, Victoria, BC, V8P 5C2, Canada}

\author[0000-0002-6221-5360]{Paul Kalas}
\affiliation{Astronomy Department, University of California, Berkeley, CA 94720, USA}
\affiliation{SETI Institute, Carl Sagan Center, 189 Bernardo Avenue, Mountain View, CA 94043, USA}
\affiliation{Institute of Astrophysics, FORTH, GR-71110 Heraklion, Greece}

\begin{abstract}
    Debris disks are dusty, optically thin structures around main sequence stars. HD~106906AB is a short-period stellar binary, host to a wide separation planet, HD~106906b, and a debris disk. Only a few known systems include a debris disk and a directly imaged planet, and HD~106906 is the only one in which the planet is exterior to the disk. The debris disk is edge-on and highly asymmetric in scattered light.  Here we resolve the disk structure at a resolution of 0\farcs38 (39\,au) with the Atacama Large Millimeter/submillimeter Array (ALMA) at a wavelength of 1.3\,mm. We model the disk with both a narrow and broad ring of material, and find that a radially broad, axisymmetric disk between radii of $\sim50-100$\,au is able to capture the structure of the observations without evidence of any asymmetry or eccentricity, other than a tentative stellocentric offset. We place stringent upper limits on both the gas and dust content of a putative circumplanetary disk.  We interpret the ALMA data in concert with scattered light observations of the inner ring and astrometric constraints on the planet's orbit, and find that the observations are consistent with a large-separation, low-eccentricity orbit for the planet.  
    A dynamical analysis indicates that the central binary can efficiently stabilize planetesimal orbits interior to $\sim$100\,au, which relaxes the constraints on eccentricity and semimajor axis somewhat.
    The observational constraints are consistent with in situ formation via gravitational instability, but cannot rule out a scattering event as the origin for HD~106906b's current orbit.
    
\end{abstract}
\section{Introduction}
Debris disks are flattened structures of dust orbiting their host stars, analogous to the Kuiper belt in our own solar system. Because the dust in debris disks is continuously replenished through collisions of planetesimals, the detection of a debris disk indicates successful formation of bodies at least 100s to 1000s of km in size \citep{wyatt08,matthews14,hughes18}. As systems evolve, planets imprint their presence on disk material via gravitational interactions in the forms of gaps \citep[e.g.,][]{su09,marino18,nederlander21}, warps \citep[e.g.,][]{mouillet97}, and spiral arms \citep[e.g.,][]{konishi16,monnier19}. Planetesimal collisions produce grains at a variety of sizes \citep{dohnanyi69}. While small grains can be affected by radiation pressure and interaction with the ISM, larger grains, imaged at millimeter wavelengths, are less susceptible to these forces, making them a more reliable tracer of a system's dynamical history \citep{wyatt06,thebault09}.  Observations of this kind of substructure in resolved disks have also been used as indirect evidence of unseen planets. In particular, a number of disks show a departure from axisymmetry in the form of eccentricity \citep[e.g.,][]{kalas05,rodigas15,sai15}. The underlying cause of eccentric debris disks is hard to pinpoint, with proposed causes including ISM interactions, stellar flybys, and interactions with eccentric planets.

HD~106906 is a 13\,Myr-old short-period binary comprising two F-type stars located in the Lower Centaurus Crux region of the Scorpius-Centaurus OB association 103.3 pc from the Sun \citep{rodet17, gaiadr2}. HD~106906 is host to a debris disk and a wide-separation planet, HD~106906b. The debris disk was first discovered as a strong infrared excess \citep{chen05} and later detected directly in scattered light with the Gemini Planet Imager (GPI), the {\it Hubble Space Telescope} (\textit{HST}), and SPHERE \citep{kalas15, lagrange16, crotts21}. The {\it HST} observations show that the NW side of the disk halo extends up to 550\,au, while the eastern component of the disk is more vertically diffuse, and extends to only 370\,au \citep{kalas15}. Conversely, GPI observations indicate that the SE component interior to $\sim$100 au is brighter than the NW component. The GPI observations also show an offset between the location of the binary and the center of the disk corresponding to an eccentricity of $\gtrsim 0.16$. Finally, the GPI observations yield a vertical FWHM of $\sim0\farcs15$ and show no evidence of asymmetry in vertical scale height, although they do indicate asymmetry in vertical offset \citep{crotts21}. The disk is nearly edge on, with inclination $\sim 85^\circ$ \citep{kalas15}.

The companion, HD~106906b, is $\sim11.9^{+1.7}_{-0.8}$\,M$_{\text{Jup}}$ \citep{daemgen17} and was directly imaged at a projected separation of 737\,au \citep{bailey14} and inclination of $\sim21 ^\circ$ from the disk midplane \citep{kalas15}. Its inferred mass is close to the deuterium-burning limit, making it unclear whether it should be considered a planet or a brown dwarf; it is therefore desirable to understand whether its formation was top-down or bottom-up so that it can be better classified. {\it HST} observations over a baseline of 14 years confirm common proper motion with the star and indicate that the planet's orbit is both potentially eccentric ($0.44\pm\substack{0.28\\0.31}$) and significantly inclined (40$^\circ\pm\substack{27\\14}$) relative to the disk midplane \citep{nguyen21}. They also show that the planet is extremely red and find tentative evidence for resolved structure surrounding it, both of which could indicate a circumplanetary disk.

The asymmetry of the disk at optical wavelengths along with the presence of a planetary companion external to the disk raises questions about whether HD~106906b may be responsible for the asymmetry. Previous dynamical studies indicated that secular gravitational interaction with an inclined, eccentric planet could create asymmetries in disk material that are qualitatively similar to the disk morphology observed in scattered light \citep{jilkova15, rodet17, nesvold17}. 

At the same time, the origin of HD~106906b is mysterious due to its high separation and the inclination of its orbit relative to the disk midplane. 
Most planets are understood to form via core accretion in a primordial gaseous disk \citep{pollack1996}, while disk fragmentation via gravitational instability has been proposed as an alternative mechanism for some planets \citep[e.g.,][]{boss1997}. Forming a giant planet more than 700\,au from its host star is unlikely via core accretion, since core formation timescales at this distance are long compared to the lifetimes of primordial disks \citep{dodson09,rafikov11}. However, mechanisms such as pebble accretion may allow core accretion far from the central star \citep[i.e.,][although whether this mechanism could operate at such an extreme separation as 700\,au is unclear]{lambrechts12, piso15}. In the disk fragmentation scenario, fragments of the disk cool very efficiently and collapse into substellar companions. This mechanism is unlikely to operate in the inner parts of protostellar disks \citep{boley2009, johnson2013}, but may play a significant role in the outer parts of extended protostellar disks \citep{stamatellos2009, vorobyov2010}.

Previous studies have also considered a scenario in which HD~106906b formed much closer to the star and was then ejected via gravitational interaction with disk material and the binary. This scenario would require some event to return the planet to a stable orbit, such as a stellar flyby. \cite{derosa19} identify candidate stars for this interaction, but dynamical modelling suggests that the approaches by those objects were most likely not close enough to result in the observed orbit for HD~106906 \citep{rodet19}. Free-floating planets present another intriguing potential source of dynamical influence, and such an interaction may be  an order of magnitude more likely than a stellar fly-by (Moore et al.~submitted).  If the planet formed within the disk and was ejected, it may also have perturbed the disk on its scattered path and produced the observed optical wavelength asymmetry via short-term gravitational interactions \citep{rodet17}.
\cite{kalas15} also presented tentative evidence from optical and near-infrared imaging data for circumplanetary dust grains, which may indicate in situ formation. However, a broader body of work on PMCs has shown that scattering is probably not the dominant scenario by which wide-separation companions are formed \citep[e.g.,][]{bryan2016,pearce2019,swastik2021}.

Here we present new ALMA observations of the system (Section \ref{observations}). In Section \ref{results} we measure basic features of the disk emission like the stellar offset and flux of the ansae.  We also present new upper limits on circumplanetary material and on the gas mass in the disk. In Section \ref{analysis} we characterize the morphology of the disk's continuum emission and analyze the planet's orbit. We also dynamically simulate interactions between the disk, the planet and the central binary, and compare our analysis with previous simulations of the disk and other dynamical criteria (Section \ref{discussion}). We summarize our conclusions in Section \ref{summary}.

\section{Observations}\label{observations}
We observed the HD~106906 debris disk with ALMA in five scheduling blocks between January and September 2018 (ALMA Project 2017.1.00979.S, PI Hughes).  Four different antenna configurations with baselines varying from 14 to 2516 m are included in the data. There are four spectral windows, each 1.875 GHz wide to provide maximum sensitivity. The three continuum windows were centered on frequencies of 228.5 GHz, 215.0 GHz, and 213.0 GHz, with channel spacings of 15.6 MHz. One spectral window was centered on the rest frequency of the CO(2-1) molecular line (230.538 GHz) with a channel spacing of 976 kHz. Table \ref{tab:obs} lists the dates, times, number of antennas, baseline lengths, time on source, synthesized beam size, and rms noise values for each scheduling block. The quasar J1107-4449 was used as the bandpass and flux calibrator for all observation blocks. J1155-5730 was used as the phase calibrator for the September track, and J1206-6138 was used as the phase calibrator for all other tracks.

Calibration, reduction, and imaging were carried out using  the Common Astronomy Software Applications (\texttt{CASA}) package \citep{mcmullin07}. Statistical weights for each visibility were calculated from the variance of nearby visibilities in the uv plane, as described in \cite{flaherty17}. 

\begin{table}
    \centering
    \caption{ALMA observations of HD~106906}
    \resizebox{\linewidth}{!}{
    \begin{tabular}{ccccccccc}
         \hline
        Date/Time (UT) & $\#$ Antennas & Baseline Lengths  & On-Source time & PWV & Beam Major Axis  & Beam Minor Axis & Beam PA  & rms noise \\
        && (m) & (min) & (mm) & (") & (") & ($^\circ$) & ($\mu$Jy beam$^{-1}$)\\
        \hline
        Jan9/10:07 & 45 & 14-2516 &48.9&2.13-2.23&0.260& 0.238 & 68.6 & 16.7\\
        Jan19/10:01 & 44 & 14-1398 &49.0&2.19-2.31& 0.431 &0.330 &52.5 & 12.1\\
        Jan24/8:14 & 43 & 15-1398 &49.0& 1.98-2.85&0.392 & 0.338 & 26.5 & 17.1\\
        Mar10/7:45 & 40 & 14-1241 &49.1&0.80-1.00 &0.527 & 0.422 & 69.0 & 19.0\\
        Sep11/19:08 & 43 & 15 -1231&49.0& 0.78-0.91&0.535 & 0.353 & 64.0 & 13.9\\
        Combined & ... &14-2516 &  245.0&...& 0.382&0.313&62.6&6.2
    \end{tabular}%
    }
    \label{tab:obs}
\end{table}

\section{Results} \label{results}
\subsection{1.3\,mm Dust Continuum}
\begin{figure}
    \centering
    \includegraphics[width=7in]{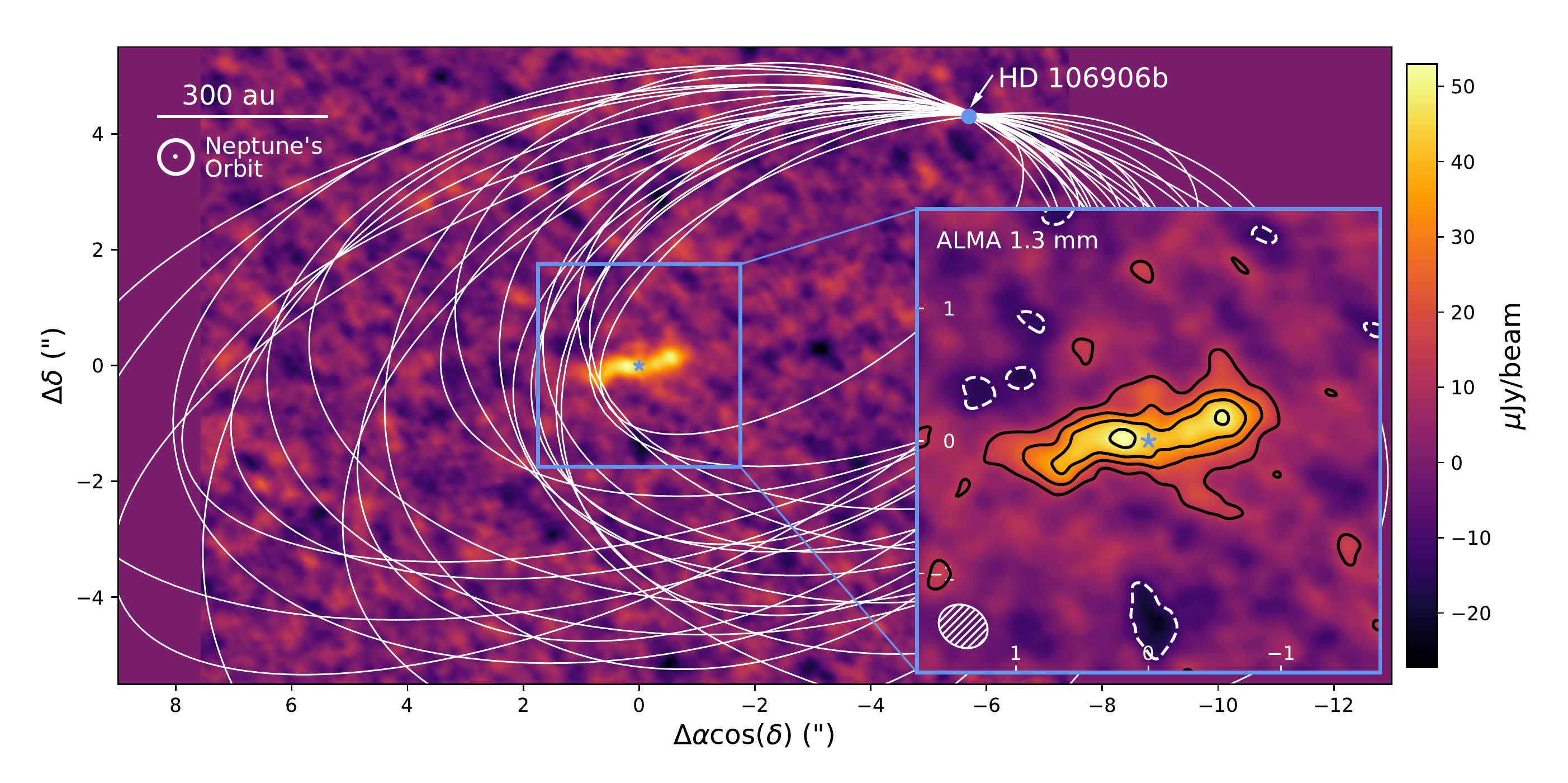}
    \caption{Naturally weighted ALMA image of the 1.3 mm continuum emission from the disk around HD~106906. The star symbol denotes the expected location of the central stellar binary, while the blue dot denotes the expected location of the planetary companion HD~106906b. There is no continuum emission visible at the location of the planet. Twenty-five orbit tracks sampled randomly from the posterior distribution of orbital elements obtained in \cite{nguyen21} are projected onto the sky plane and represented by the white ellipses. A zoomed-in image of the disk emission is inset, with contour levels at [-2,2,4,6,8] x $\sigma$ where $\sigma$ is the rms noise in the image, 6.2 $\mu$Jy beam$^{-1}$. The hatched ellipse represents the size and orientation of the synthesized beam, 0\farcs38x 0\farcs31. The size of Neptune's orbit is displayed beneath the scale bar for visual comparison.}
    \label{fig:1}
\end{figure}
The observation blocks were combined to generate an image of dust continuum emission around HD~106906. Figure \ref{fig:1} shows a naturally weighted image of the data set generated using the \texttt{CASA} task \texttt{tclean}. The figure also shows the expected positions of HD~106906 \citep{gaiadr2} and HD~106906b, as well as twenty-five orbit tracks sampled randomly from the posterior distribution of orbital elements from \cite{nguyen21}. The image shows two peaks of dust brightness, one on either side of the binary. Beyond the peaks, there is some extended flux to the southeast and northwest, with a possible asymmetry towards the southeast. The vertical structure of the disk is unresolved, as the apparent height is approximately equal to the size of the synthesized beam.

Using the \texttt{CASA viewer} task, we measure the distance between the locations of peak intensity in the SE and NW components, and find $\Delta \alpha \cos \delta = 0\farcs75 \pm 0.04$ and $ \Delta\delta = 0\farcs16 \pm 0.04$. We then calculate the offset between the midpoint of the peaks and the expected position of the star, including a proper motion correction. We find a stellar offset of $\Delta\alpha \cos \delta= 0\farcs17\pm0.06$ (18\,au $\pm 6$), $\Delta\delta = -0\farcs10 \pm 0.06$ (10\,au $\pm6$), which would correspond to an eccentricity of $e=0.5\pm0.2$ (which should be considered a lower limit due to projection effects onto the sky plane). In addition to the uncertainty due to ALMA's resolution, we also accounted for the uncertainty in the binary's position as obtained by \textit{Gaia}.  However, it is not clear a priori whether the peaks in the image are meaningful in a statistical sense; they could just be noise peaks in a low-SNR regime.  We therefore conduct a more careful visibility-domain analysis of the potential offset between disk and star in Section~\ref{analysis} below, which shows that the offset is not necessarily significant.  Our dynamical analysis in Section~\ref{sec:disk_structure} shows that even if the offset is significant, it is not necessarily due to eccentricity, but could conceivably just be caused by the reflex motion from the planet.

Eccentricity can lead to one disk anse becoming brighter than the other, an effect referred to as pericenter glow \citep[which tends to dominate at infrared wavelengths;][]{wyatt99} or apocenter glow \citep[which tends to dominate at millimeter wavelengths;][]{pan16}. At longer wavelengths, variation in the surface density of dust becomes more important than dust temperature or proximity to the central star (for scattering). It has been proposed that this would result in apocenter glow, since each particle spends more of its orbit at apocenter than pericenter. \cite{lynch21} add nuance to this idea, finding that an eccentric disk can result in either apocenter or pericenter glow at long wavelengths depending on the resolution of the observations, but that the relationship found in \cite{pan16},
$$\text{apo/peri flux ratio}\simeq \left(\frac{1-e/2}{1+e/2}\right) \left(\frac{1+e}{1-e}\right),$$
still holds in the case of low angular resolution compared to the radial width of the disk. We find an apo/peri peak flux ratio of $1.0 \pm 0.2$, corresponding to a circular disk with a 3-sigma upper limit on eccentricity of $0.4$.  

While these two methods of estimating the disk's eccentricity apparently disagree, we must consider that the location of apparent peaks at low SNR can be influenced by random fluctuations, and second, that factors other than eccentricity (like an offset between a system's center of mass and the location of the star) can potentially cause offsets between a disk's center and its host star.  We therefore follow up this preliminary image-domain analysis with a more detailed visibility-domain analysis in Section~\ref{analysis} that is less susceptible to the location of random noise peaks, as well as a dynamical analysis in Section~\ref{discussion} that examines the impact of the planet on the morphology of the star-disk system. 

\subsection{Upper Limit on the Dust Content of a Putative Circumplanetary Disk}

Circumplanetary disk detection at millimeter wavelengths has so far proven elusive.  Most searches that have taken place focused on so-called ``planetary-mass companions," or PMCs, which are typically defined by their youth \citep[most are $<100$\,Myr old;][]{bowler2016} and wide orbits \citep[$>100$\,au;][]{bowler2014}.  Even in the youngest systems, searches with ALMA have mostly returned upper limits on circumplanetary material \citep{wu2017,wu2020}, which are so low that they probably require a combination of fast radial drift and a lack of dust traps, or some sort of tidal truncation to $\lesssim10$\,au, to explain \citep{rab2019,wu2020}.  One interesting exception to this rule is the detection of a circumplanetary disk around the free-floating planetary-mass object OTS~44 \citep{bayo2017}, which was detected at a level of 101\,$\mu$Jy, very close to the typical 100-200\,$\mu$Jy rms noise level of most searches for circumplanetary disks around PMCs. Recently, circumplanetary disks have also been detected around the PMCs PDS 70c and SR 12c, with flux levels of $86\pm16$\,$\mu$Jy\,beam$^{-1}$ and $127\pm14$\,$\mu$\,Jy\,beam$^{-1}$, respectively \citep{benisty2021,wu2022}. 

\citet{kalas15} pointed out that the infrared excess of HD~106906b is second in brightness only to that of FW~Tau~b, which was the first companion to exhibit a detectable millimeter-wavelength flux indicative of the presence of a circumplanetary disk \citep{kraus2015}.  The other two known objects with infrared colors comparable to that of HD~106906, GSC 6214-210B and 1RXS~1609-2015B, both exhibit excess emission, and one of the two also displays both H$\alpha$ and Pa$\beta$ signatures indicative of accretion onto the companion object. It is therefore plausible that HD~106906b might host a circumplanetary disk. While accretion indicators place an upper limit on its accretion rate of $4.8\times10^{-10}$\,M$_\mathrm{Jup}$\,yr$^{-1}$ \citep{daemgen17}, the interpretation of accretion signatures is complex and could be higher or lower depending on adjustments to the extrapolation of accretion rate models from the stellar to substellar regime, or due to detectable H$\alpha$ emission generated by a shock where circumstellar material falls onto the circumplanetary disk \citep[e.g.][]{aoyama2018,aoyama2021}.  Our ALMA observations yield only an upper limit on its flux (similar to the majority of PMC targets, although our limits are more stringent than most surveys by a factor of $\sim3-5$).  Here we quantify the upper limit and examine its implications. 

We measured the upper limit in units of flux per beam at the expected location of the planet — ${\Delta \alpha = -5\farcs6}$, ${\Delta \delta = +4\farcs3}$ \citep{nguyen21}. We then used estimates of the planet’s mass, the star’s mass, the semi-major axis, and the eccentricity of the planet’s orbit to estimate the planet’s Hill radius. A possible circumplanetary disk must lie within the Hill radius and likely lies within one third of it \citep{quillen1998,ayliffe2009,martin2011,shabram2013}, so we used both as estimates for the circumplanetary disk’s radius. We calculated an estimate for the Hill radius  using the equation $R_\mathrm{Hill} = a(1-e)\left(\frac{m}{3M}\right)^{1/3}$, where $a$ is the semi-major axis, $e$ is the eccentricity, $m$ is the planet’s mass, and $M$ is the star’s mass.  We adopted $m = 11\,M_\text{Jup}$, $M_\star = 2.5\,M_\sun$ \citep{bailey14}, semi-major axis $a = 850$\,au and orbital eccentricity $e = 0.44$ \citep{nguyen21}, which yielded a Hill radius estimate of 53\,au. 

We next calculated the upper limit on the integrated flux within radii of 1/3\,$R_\mathrm{Hill}$ and 1\,$R_\mathrm{Hill}$. First, we measured the rms noise within a single $0\farcs38 \times 0\farcs31$ naturally weighted beam to be 6.6\,$\mu$Jy/beam.  Since the linear scale of the beam at a distance of 103.3\,pc is $39\times32$\,au, we consider the beam area to be comparable to that covered by a region of radius 1/3 R$_\mathrm{Hill}$, and multiply the rms noise by a factor of 3 to yield an upper limit of 20\,$\mu$Jy within 1/3 $R_\mathrm{Hill}$.  To calculate the upper limit within the Hill radius of 53\,au we noted that the area covered should be equivalent to approximately $N = 9$ synthesized beams, and multiplied the upper limit within one beam by $\sqrt{N} = 3$ to yield an upper limit of 60\,$\mu$Jy. 

Finally, we translated the flux upper limits into disk parameters, using two methods: one taking into account only the dust disk, and the other taking into account models of accreting circumplanetary disks from \cite{zhu2018}, who scale a simple $\alpha$-disk model of an accreting disk to examine the expected relationship between the millimeter flux and basic disk parameters like the accretion rate, viscosity parameter $\alpha$, and the mass, radius, and temperature of the central object. 

To convert our millimeter flux limits to dust masses, we use Equation 1 from \citet{andrews2005}:
\begin{equation}
M_d = \frac{d^2 F_\nu}{\kappa_\nu B_\nu(T_c)}
\end{equation}
where $d$ is the distance to the source, $F_\nu$ is the flux density (in this case an upper limit), $B_\nu$ is the Planck function at a characteristic temperature $T_c$, and $\kappa_\nu$ is the dust mass opacity (ignoring gas), which we assume to be
\begin{equation}
    \kappa_\nu = 10 \left(\frac{\nu}{\rm{10^{12}} \ Hz}\right)^\beta \ \rm{cm^2 \ g^{-1}}
    \label{eq:opacity}
\end{equation}
where $\nu$ is the frequency of observation and we adopt a spectral index $\beta=1$ \citep{beckwith1991}.  To estimate an appropriate characteristic temperature for the circumplanetary dust disk, we scale a value of $\sim$20\,K (appropriate for a 1\,M$_\sun$ star with an age of 1\,Myr) by a factor L$_*^{1/4}$, using a value of $2.3\times10^{-4}$\,L$_\sun$ for the luminosity of HD~106906b \citep{bailey14}.  This calculation yields an estimated characteristic temperature of $\sim3$\,K (which is about as low as it could reasonably be, thereby providing a conservative upper limit), which means that the upper limits on the flux density translate to upper limits on the dust mass of $0.32$\,M$_\earth$ and $0.96$\,M$_\earth$ within 1/3 and 1\,R$_\mathrm{Hill}$, respectively.   

For purposes of comparison to the accreting circumplanetary disk models from \citet{zhu2018}, we scaled both flux density values to a distance of 140\,pc, yielding upper limits of 11 and 33\,$\mu$\,Jy\,beam$^{-1}$ within 1/3 and 1 $R_\mathrm{Hill}$, respectively. We then compared these flux values with the expectations for an accreting circumplanetary disk by treating them as viscous (rather than irradiated) circumplanetary disks and interpolating Table 1 from \citet{zhu2018}.  The result is Figure 2, which illustrates how the upper limits from our 1.3\,mm ALMA observations compare with the parameter space.  Each of the four panels shows a different product of planetary mass and accretion rate ($M_p \dot{M_p}$), and the axes show the parameter space defined by the combination of viscosity parameter $\alpha$ (on the x-axis) and outer radius (on the y-axis).  The solid orange line shows the flux density upper limit within a radius of 1\,$R_\mathrm{Hill}$ and the dotted line shows the flux density upper limit within a radius of 1/3\,$R_\mathrm{Hill}$.  The regions of parameter space with higher flux (brighter colors) than the lines are ruled out by our observations, whereas the regions to the lower right are consistent with our observations.  Essentially, our observations rule out high-mass planets with high accretion rates (the two left panels), but do not rule out the presence of even a large disk around a low-mass planet with a large viscosity parameter.  For example, given the nominal planet mass of 11\,M$_\mathrm{Jup}$ and the accretion upper limit of $4.8\times10^{-10}$\,M$_\mathrm{Jup}$ from \citet{daemgen17}, a disk of essentially any radius with a viscosity parameter of $10^{-1}-10^{-2}$ would still be consistent with our observations. 

\begin{figure}
    \centering
    \includegraphics[scale=0.8]{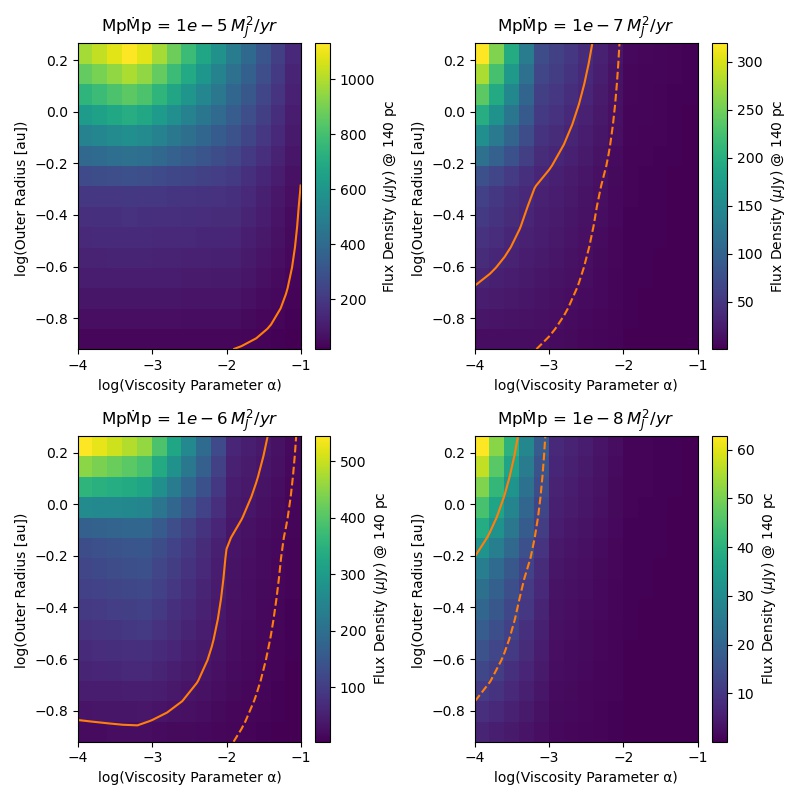}
    \caption{Interpolated model of accreting circumplanetary disks from \citet{zhu2018} with viscocity parameter $\alpha$ on the x-axis, outer radius on the y-axis, and flux density in $\mu$\,Jy as color, adjusted to a distance of 140\,pc. The dashed contour line represents the upper limit of 11\,$\mu$\,Jy\,beam$^{-1}$ within a radius of 1/3\,$R_\mathrm{Hill}$, while the solid line represents the upper limit of 33\,$\mu$\,Jy\,beam$^{-1}$ within a radius of 1\,$R_\mathrm{Hill}$.}
    \label{fig:my_label}
\end{figure}

\subsection{CO(2-1) Upper Limits}

We searched the ALMA data for gas near the position of the HD~106906 debris disk as well as the position of the HD~106906b companion. Using the rest frequency of CO as 230.53800 GHz taken from Splatalogue, we used the CASA command \texttt{uvcontsub} to subtract continuum emission, excluding channels near the expected frequency of the line and at the edges of the bandpass. We used the \texttt{cvel} task to convert the reported velocities from the heliocentric to the Local Standard of Rest (LSR). We then imaged the data using \texttt{tclean} and produced moment maps using \texttt{immoments}, integrating channels within $\pm10.5$\,km\,s$^{-1}$ of the expected LSR  velocity of the star. 

The heliocentric velocity of the HD~106906 system, $12.18 \pm 0.15$\,km\,s$^{-1}$, was taken rom \citet{derosa19}. We converted it to a kinematic LSR velocity of 4.38\,km\,s$^{-1}$ using publicly available conversion code\footnote{developed for use in \cite{tollerud2016} and available at https://github.com/eteq/erikutils/blob/master/erikutils/velocities.py}.  We then estimated a 3$\sigma$ upper limit on the integrated CO flux for the circumstellar disk, the area within one Hill radius of the companion, and the area within 1/3 Hill radius of the companion.  We multiplied 3 times the rms noise in the moment zero image of 3.0\,mJy\,km\,s$^{-1}$ by the square root of the number of beams within the relevant area ($\sqrt{N_\mathrm{beams}}$), which we estimated for the debris disk around HD~106906 by calculating the area enclosed within the 2$\sigma$ flux contours of the continuum emission.  We converted the resulting flux into mass upper limits of CO gas using the following relationship, which uses the expected level populations $X_u$ based on the assumed excitation temperature of the gas $T_{ex}$ along with the Einstein coefficient $A$ to translate between the observed flux and the mass of CO molecules in the disk:

\begin{equation}
\label{eqn:Mass Conversion Equation}
M=\frac{4\pi}{h\nu_{0}}\frac{F_{Ju-Jl}m_{mol}d^2}{A_{Ju-Jl}X_{u}}
\end{equation}
\begin{equation}
\label{eqn:Upper State Ratio}
X_u=\frac{g_u}{Q(T)}e^{{-E_u}/{kT_{ex}}}
\end{equation}
\noindent
where $F_{Ju-Jl}$ is the flux upper limit and d is the distance to the source. $m_{mol}$ is the molecular mass and $A_{Ju-Ji}$ is the Einstein A coefficient of the CO(2-1) transition. $X_u$ is the fraction of molecules in the upper energy state, and this is calculated through $g_u$, the degeneracy of the upper state. $Q(T)$ is the tabulated partition function at $T=37.5K$, and $E_u$ is the energy of the upper state. All of the above quantities are taken from the Cologne Database for Molecular Spectroscopy \citep[CDMS;][]{endres2016}. $T_{ex}$ is the excitation temperature, for which we assume the same 31\,K as for the dust temperature.  

Table~\ref{tab:gas_limits} presents the beam areas and flux and mass upper limits for the debris disk around HD~106906AB as well as the region enclosed within an area with radii of 1/3 and 1 Hill radius around the companion, HD~106906b.  The mass upper limits are for CO only; the total gas mass could be substantially larger if the gas were either dominated by H$_2$ or photodissociated into atomic C and O.  

\begin{deluxetable}{cccc}
\centering
\tablewidth{0pt}  
\tablecaption{CO(2-1) Upper Limits \label{tab:gas_limits}
}
\tablehead{
  \colhead{Region} & \colhead{N$_\mathrm{beams}$} &
 \colhead{Flux} &  \colhead{Mass} \\
\colhead{} & \colhead{} &\colhead{(Jy\,km\,s$^{-1}$)} &\colhead{(M$_{\Earth}$)}}
\startdata
Debris Disk  & 9.3 & $<2.7 \times 10^{-2}$ & $<9.3 \times 10^{-6}$\\
Hill Radius & 9.0 & $<2.7 \times 10^{-2}$ & $<9.3 \times 10^{-6}$\\ 1/3 Hill Radius & 1.0 & $<8.9 \times 10^{-3}$ & $<3.1 \times 10^{-6}$\\
\enddata
\end{deluxetable}

\section{Analysis} \label{analysis}
\subsection{Parametric Disk Model}

To better determine the distribution of dust in the system, we model the continuum observations in the visibility domain using a parametric model. We generated synthetic model images of disks with a variety of geometries using the \texttt{Galario} function \texttt{chi2Profile} \citep{tazzari18}, which calculates synthetic visibilities from a radial brightness profile and then calculates a $\chi^2$ metric by comparing the synthetic visibilities with the data in the uv plane. Comparing data in the uv plane is preferable to the image domain, as uncertainties are well-characterized and the choice of imaging parameters does not affect the comparison between data and model. We fit the models to the data using an affine-invariant MCMC sampler \citep{goodman10} implemented using the software package \texttt{emcee} \citep{foremanmackey13}. The goodness of fit was evaluated by a log-likelihood metric, $\ln \mathcal{L} = -\chi^2/2$. The MCMC code uses an ensemble of walkers to explore the parameter space, with each walker moving (or not) according to the probability that a new walker position provides a better fit than the previous walker position. After initial burn-in (described below), the process results in a set of model parameters that sample the posterior probability distribution for each parameter.

We performed several MCMC runs with a variety of disk geometries. We first modeled the disk as a Gaussian ring, with radial brightness profile
$$F(R) = F_0e^{-\left(\frac{R-R_0}{2\sigma}\right)^2}$$
where $F_0$ is the peak flux of the ring, $R_0$ is the radius of that peak flux, and $\sigma$ is the standard deviation, representing the width of the ring. Initially, we considered a narrow ring with a fixed standard deviation of 0\farcs.01 ($\sim 1$\,au), corresponding to a full width at half maximum (FWHM) lower than the angular resolution of the observations. We varied six parameters: $R_0$, $F_0$, the position angle of the disk major axis (PA), the inclination of the disk relative to the observer's line of sight ($i$), and the position offset of the disk in right ascension ($\Delta \alpha$) and declination ($\Delta\delta$) relative to the pointing center of the observation. All parameters were sampled linearly except for the brightness, which was sampled logarithmically, equivalent to using a log-uniform prior. The initial run revealed two issues that indicated a narrow ring might not capture the structure of the disk. First, significant ($>3\sigma$) residuals remained in the eastern extension of the disk after subtracting the best fit model. Second, the best fit and median total fluxes found by the algorithm were lower than the total flux derived from imaging the observations (0.16$\pm \substack{0.03\\0.03}$ mJy for the median ring model, versus 0.22 mJy for the data image). This is because a significant portion of the disk's emission is not represented in the model.

In order to resolve these problems, we then attempted to model the extended flux of the disk. To remain agnostic about the functional form of the extended structure, we explored both a similarity solution model \citep{lynden74} and a broad Gaussian ring. We used a functional form for the similarity solution given by 
$$F(r) = F_0 \frac{R}{R_c}^{\gamma}e^{\left(\frac{R}{R_c}\right)^{\gamma+2}}$$
where $F_0$ is the peak flux of the disk, $R_c$ is the critical radius, and $\gamma$ is the power law index for the inner edge of the disk. The broad Gaussian ring model is identical to the narrow ring model, but we allowed the value of $\sigma$ to vary rather than fixing it.

These two functional forms produced very similar brightness profiles. Both yielded noise-like residuals, even though the model images did not show the same double-peaked structure visible in the data image. The best fit Gaussian and similarity solutions (which have the same number of free parameters) had an insignificant $\Delta \chi^2$, 0.2, indicating that they fit the data visibilities equally well. Since the similarity solution and Gaussian ring models both included one more free parameter than the narrow ring model, we expect them to provide a better fit simply due to the number of degrees of freedom. In order to penalize the change of degrees of freedom when evaluating $\Delta \chi^2$, we employ the AIC, a form of the Aikake Information Criterion, and the BIC, a form of the Bayesian Information Criterion. The BIC penalizes additional degrees of freedom more than the AIC, and also considers the sample size of the observations. When calculated, the AIC test slightly prefers the Gaussian ring to the narrow ring, with a $\Delta$ AIC of 3.8. In contrast, the BIC for the narrow ring is lower than that Gaussian ring, with a $\Delta$ BIC  of 9.9. Although the Gaussian ring yields a lower $\chi^2$, the AIC and BIC disagree over whether this difference warrants the addition of a parameter, so one model is not significantly better than the other. For the Gaussian ring, we see a slight degeneracy between $R_0$ and the FWHM of the disk, most likely because the inner edge is not resolved so both parameters are responding primarily to the location of the disk's outer edge.

\begin{table}[]
    \centering
    \caption{MCMC Priors}
    \resizebox{\linewidth}{!}{
    \begin{tabular}{ccc|cccc}
     \hline
        Parameter & Similarity Solution & Similarity Solution&Parameter& Gaussian Ring & Narrow Ring  & Gaussian Ring  \\
        &&+ Narrow Ring&&& &  + Narrow Ring\\
        \hline
        $R_c (")$           & [0,4]         & [0,4]         & $R_0 (")$           & [0.01,5]      & [0.01,5]      & [0,2.5]\\
        $F_0$ (Jy/sr)       & [1,$10^{10}$] & [1,$10^{10}$] & $F_0 (Jy/sr)$       & [1,$10^{10}$] & [1,$10^{10}$] & [1,$10^{10}$]\\
        PA ($^\circ$)       & [0,180]       & [0,180]       & PA ($^\circ$)       & [0,180]       & [0,180]       & [0,180]\\
        $i$ ($^\circ$)      & [0,90]        & [0,90]        & $i$ ($^\circ$)      & [0,90]        & [0,90]        & [0,90]\\
        $\Delta\alpha$(") & [-1,1]        & [-1,1]        & $\Delta\alpha$(") & [-1,1]        & [-1,1]        & [-1,1]\\
        $\Delta\delta$(") & [-1,1]        & [-1,1]        & $\Delta\delta$(") & [-1,1]        & [-0.5,0.5]    & [-0.5,0.5]\\
        $\gamma$            &[-5,5]         & [-5,5]        & $\sigma$            & [0.01,5]      &        0.01       & [0.01,2.5]\\
        $R_{0,comp}(")$            &               & [0.01,5]      &  $R_{0,comp} (")$               &               &               & [0.01,2.5]\\
        $F_{0,comp}$ (Jy/sr)       &               & [1,$10^{10}$] & $F_{0,comp}$ (Jy/sr)       &               &               & [1,$10^{10}$]\\
        $\Delta\alpha_{comp}$ (") &              & [-1,1]        & $\Delta\alpha_{comp}$ (")&               &               &[-1,1]\\
        $\Delta\delta_{comp}$(")  &              & [-0.5,0.5]    & $\Delta\delta_{comp}$ (")&               &               &[-0.5,0.5]\\
        \hline
    \end{tabular}
    }
    \label{tab:mcmc_priors}
\end{table}

Finally, we attempted to capture both the extended and compact flux in a single model. To do so, we considered a narrow ring in combination with either a similarity solution or Gaussian ring. For each of these models, the compact structure was allowed a separate peak flux ($F_{0, comp}$), radius of peak flux ($R_{0,comp}$), and set of position offsets ($\Delta\alpha_{comp}, \Delta\delta_{comp}$) from the extended structure, but shared their position angle and inclination.
The similarity solution with the narrow ring resulted in a marginally better fit than the Gaussian ring with the narrow ring, $\Delta\chi^2 = 2.2$. Since the addition of the narrow ring adds four additional parameters, the AIC prefers the similarity solution without the ring with a $\Delta$ AIC = 5.2. In addition, the total ring brightness found by the MCMC algorithm was very low compared to the total brightness of the extended structure. It was also not statistically significant, so we were only able to derive an upper limit on the flux of the ring, which further suggests that the extended flux component adequately describes the emission morphology in the ALMA images. As a result, the best fit parameters for the models containing both a narrow ring and extended structure are not representative of the real morphology of the disk.

\begin{table}[]
    \centering 
    \caption{MCMC Fitting Results - Similarity Solution, Broad Ring, Narrow Ring}
    \begin{tabular}{ccc|ccccc}
    \\
    \hline
    \multicolumn{1}{c}{Parameter} & \multicolumn{2}{c}{Similarity Solution}& \multicolumn{1}{c}{Parameter} & \multicolumn{2}{c}{Gaussian Ring} & \multicolumn{2}{c}{Ring}\\
    \cline{2-3}
    \cline{5-8}
     &Best Fit & Median & & Best Fit & Median & Best Fit & Median\\
     \hline
    $R_c$ (")              & 1.0  & 0.9 $\pm \substack{0.3 \\ 0.2}$  & $R_0$ (")        & 0.6  & $<$ 1.1$^a$   & 0.5  & 0.6 $\pm\substack{0.2\\0.1}$\\
    Flux (mJy)             & 0.23 & 0.22 $\pm\substack{0.04\\ 0.04}$ & Flux (mJy)       & 0.23 & 0.22 $\pm \substack{0.04\\0.03}$ & 0.19 & 0.16 $\pm \substack{0.03\\0.03}$\\
    PA $(^\circ)$          & 102  & 103 $\pm \substack{5\\5}$        & PA $(^\circ)$    & 102  & 103 $\pm \substack{4\\4}$       & 99   & 100 $\pm \substack{6 \\7}$\\
    $i$ $ (^\circ)$        & 84   &   $>43^a$                        & $i$ $ (^\circ)$  & 84   & $>59^a$         & 70   & 73 $\pm \substack{6\\6}$\\
    $\Delta\alpha (")$     & 0.0 & 0.0 $\pm \substack{0.1\\0.1}$     &$\Delta\alpha (")$& 0.0  & 0.0 $\pm \substack{0.1\\0.1}$   & -0.1 & -0.1 $\pm \substack{0.2\\0.1}$\\
    $\Delta\delta (")$     & 0.00 & 0.01 $\pm \substack{0.05\\0.05}$ &$\Delta\delta (")$& -0.01 & 0.00 $\pm \substack{0.05\\0.05}$ & 0.04 & 0.04 $\pm \substack{0.06\\0.09}$\\
    $\gamma$               & 0.6 & 0.9 $\pm \substack{1.9\\0.9}$     & FWHM (")         & 0.8  & 1.0 $\pm \substack{0.4\\0.4}$   & 0.02355$^b$ & 0.02355$^b$\\
    Ln prob                & -3400147.3 &                            &                  & -3400147.2 &        & -3400150.1\\
    \hline
    \end{tabular}
    \vspace{.4cm}
    
     NOTE -- $^a$ The lower limits on inclination of 43$^\circ$ and 59$^\circ$ represent the 0.3rd percentile of the posterior distribution, while the upper limit on $R_0$ of $1 \farcs 1$ represents the 99.7th percentile of the posterior distribution.
     
     $^b$ The FWHM of the narrow ring was fixed at a value lower than the angular resolution of the observations.
    \label{tab:mcmc1}

\end{table}

\begin{figure}
    \centering
    \includegraphics[scale=0.6]{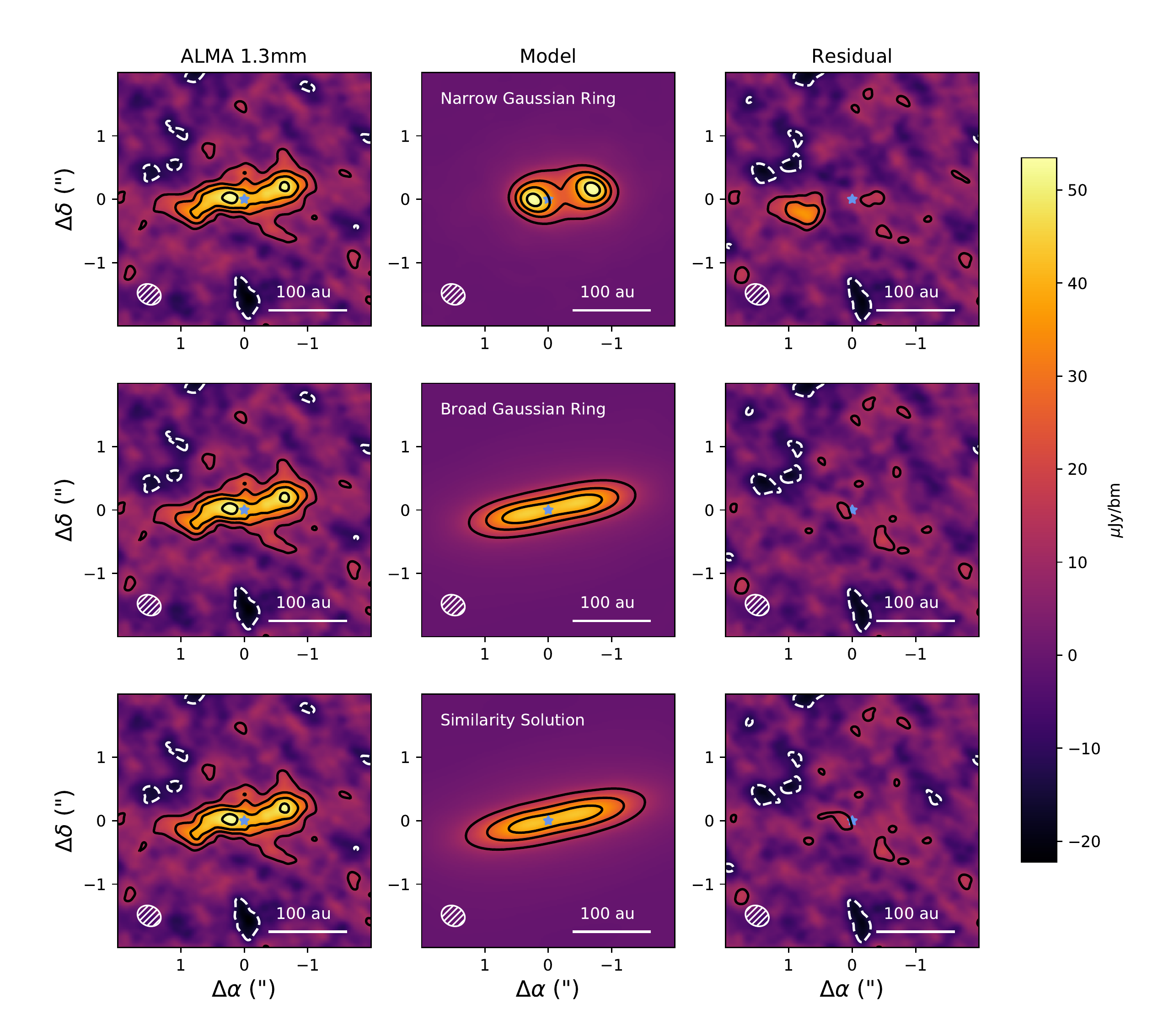}
    \caption{(Left column) Naturally weighted ALMA image of the 1.3 mm continuum emission from the HD~106906 system. (Center column) Best fit models for each functional form sampled at the same baseline lengths and orientations as the ALMA data. (Right column) Residual images after subtracting the models from the data in the visibility domain. The top row shows the narrow Gaussian ring model, the middle row shows the broad Gaussian ring model, and the bottom row shows the similarity solution model. Contour levels and symbols are as in Figure $\ref{fig:1}$.}
    \label{fig:dmr}
\end{figure}

The limits on priors for all parameters are listed in Table \ref{tab:mcmc_priors}.
For all models, we used 50 walkers and ran the chain for 2000 steps. In all cases, the burn-in period was estimated by eye based on where the lnprob values seemed to reach a consistent maximum, but had typical values of 250 to 300 steps. We also performed an autocorrelation analysis, which showed that while the autocorrelation time had not leveled off by the end of each chain, all of the parameters had stabilized such that the fractional error in the mean was a few percent or less. Figure \ref{fig:dmr} shows the data image (left) compared with the best-fit model (center), sampled at the same baseline separations and orientations and imaged with the same parameters as the data, and the residuals (right) for the narrow ring, Gaussian ring, and similarity solution models. The lack of significant residuals for the Gaussian ring and similarity solution indicate that they adequately fit the data.

In order to derive an upper limit on the brightness asymmetry of the two peaks, we also ran a model with a point source (a two-dimensional Gaussian with FWHM much smaller than the synthesized beam) at the location of each of the two ansae in the best fit narrow ring model. We fixed the position of the points and only fit the brightness of each point, allowing them to vary independently. From the posteriors, we derive an apo/peri center flux ratio of $1.0 \pm \substack{0.3\\0.3}$, corresponding to an 3$\sigma$ upper limit on eccentricity of 0.6 using the relationship in \cite{pan16}. This agrees with the limits on the apo/peri flux ratio and eccentricity we found in our image domain analysis, albeit with a slightly higher-eccentricity upper limit.

\begin{figure}
    \centering
    \includegraphics[scale=0.5]{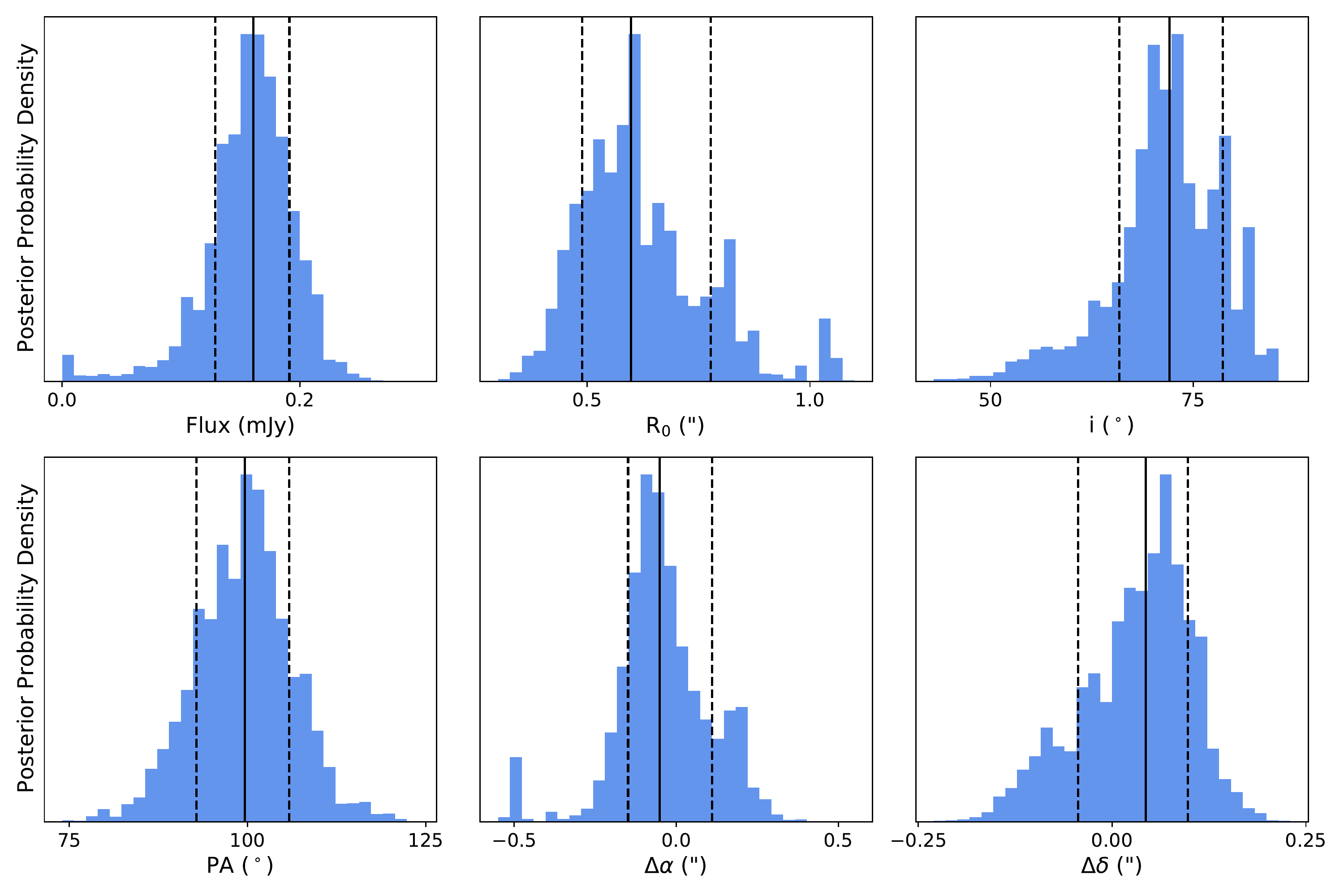}
    \caption{Histograms of the posterior probability distributions for the narrow ring model. The central  line designates the median of each distribution while the outer dashed lines mark the 16th and 84th percentiles.}
    \label{fig:ring_hist}
\end{figure}

\begin{figure}
    \centering
    \includegraphics[scale=0.5]{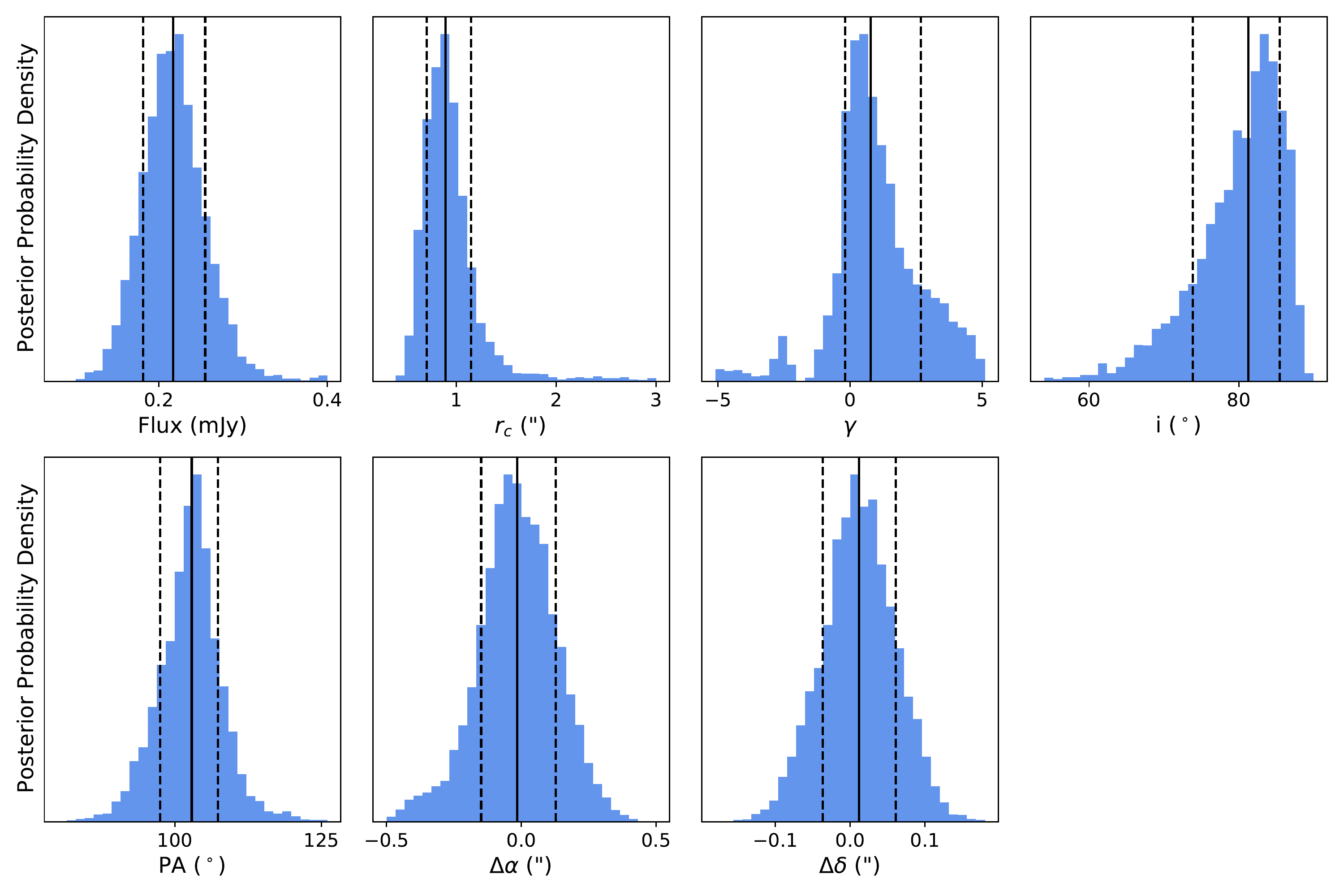}
    \caption{Histograms of the posterior probability distributions for the similarity solution model. The central  line designates the median of each distribution while the outer dashed lines mark the 16th and 84th percentiles.}
    \label{fig:SS_hist}
\end{figure}

\begin{figure}
    \centering
    \includegraphics[scale=0.5]{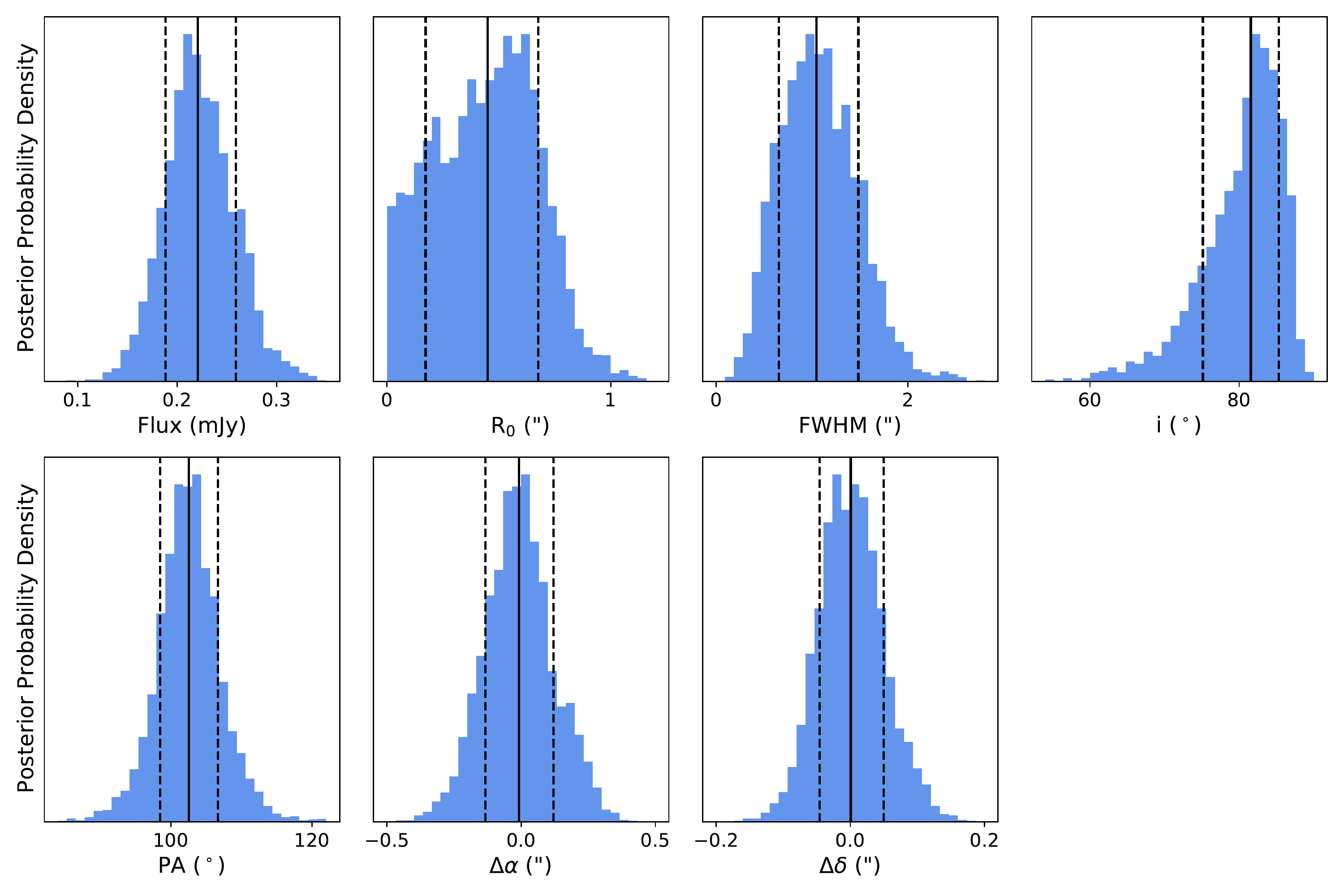}
    \caption{Histograms of the posterior probability distributions for the Gaussian ring model. The central  line designates the median of each distribution while the outer dashed lines mark the 16th and 84th percentiles. }
    \label{fig:gausian_hist}
\end{figure}

Tables $\ref{tab:mcmc1}$ and $\ref{tab:mcmc2}$ present the best-fit models, as well as the median and uncertainties of each parameter, given by the 16th and 84th percentiles of the posterior distribution, for all functional forms. We apply algebraic transformations to the posteriors to derive the total flux from the functional form and to derive the FWHM of the Gaussian rings from $\sigma$. We also calculate the position offset of the model from the location of the stellar binary as constrained by \cite{gaiadr2} and include that in the table rather than the offset from the observation's pointing center. We ignore the star's position uncertainty when producing the posterior probability distribution, as it is small compared to the standard deviation of the distribution, 0\farcs03 for both right ascension and declination. Figures $\ref{fig:ring_hist}$, $\ref{fig:SS_hist}$ and $\ref{fig:gausian_hist}$ show the probability distribution of these posteriors, with the same manipulations.

The results of our MCMC fits demonstrate that the HD~106906 disk emission is well described by a broad underlying flux distribution extending from the center of the disk to $100 \pm 20$ au (calculated as r + HWHM for the Gaussian ring). The inner radius is not well resolved, although the diffuse nature of the emission (as opposed to flux being concentrated in the central beam) suggests consistency with an inner radius of around 50\,au as suggested by \citet{kalas15}.  There are no statistically significant ($> 3\sigma$) residuals, which indicates that an azimuthally symmetric and circular distribution of flux represents the data well. Fitting the two sides of the disk independently yields an upper limit on eccentricity of 0.6. In addition, the models do not require a significant offset between the center of the disk and the location of the stellar binary. A narrow ring is able to represent the data about as well as an extended flux distribution according to the AIC and BIC, although the best fit narrow ring model does result in $> 3\sigma$ residuals, unlike the extended-disk models, which suggests a slight preference for an extended flux distribution.

\begin{table}[]
    \centering
    \caption{MCMC Fitting Results - Combination of Narrow Ring and Extended Flux Distribution}
    \begin{tabular}{ccc|ccc}
    \multicolumn{1}{c}{Parameter} & \multicolumn{2}{c}{Similarity Solution + Narrow Ring} &\multicolumn{1}{c}{Parameter}& \multicolumn{2}{c}{Gaussian Ring + Narrow Ring}\\
    \cline{2-3}\cline{5-6}
     &Best Fit & Median& & Best Fit & Median \\
     \hline
    R$_c$ (")            & 0.9      &  0.9 $\pm \substack{0.3\\ 0.2}$  & R$_0$ (")                & 0.5   & $<2.4^a$\\
    Flux (mJy)           & 0.21     & 0.22 $\pm\substack{0.04\\0.04}$  & Flux (mJy)               & 0.23  & 0.21 $\pm \substack{0.04\\0.06}$\\
    $\Delta\alpha (")$   & 0.6      & 0.0  $\pm \substack{0.7\\0.6}$   & $\Delta\alpha (")$  & 0.0   & 0.0 $\pm \substack{0.2\\0.2}$\\
    $\Delta\delta (")$   & -0.3     & 0.0  $\pm \substack{0.6\\0.6}$   & $\Delta\delta (")$ & 0.00  & 0.01 $\pm \substack{0.06\\0.05}$\\
    $\gamma$             & 1.9      & 0.3  $\pm \substack{1.9\\0.9}$   & FWHM (")             & 1.0   & 1.1 $\pm \substack{0.6\\0.4}$\\
    $R_{0, comp} (")$    & 0.4      & 1.7  $\pm\substack{1.7\\1.2}$    & $R_{0, comp} (")$             & 1.5   & 0.8 $\pm\substack{1.0\\0.4}$ \\
    Flux$_{comp}$ (mJy)  & 0.02 & $<0.12^a$                        & Flux$_{comp}$ (mJy)          & 2 $*10^{-4}$ & $<0.23^a$ \\
    $\Delta\alpha_{comp} (")$ & 0.0     & -0.2 $\pm\substack{0.5\\0.5}$    & $\Delta\alpha_{comp} (")$ & -0.1  & 0.0 $\pm\substack{0.7\\0.6}$ \\
    $\Delta\delta_{comp} (")$ & 0.0     & -0.1 $\pm\substack{0.3\\0.3}$    & $\Delta\delta_{comp} (")$ & 0.5   & 0.0  $\pm \substack{0.3\\0.3}$\\
    PA $(^\circ)$        & 104      & 103 $\pm \substack{5\\5}$        & PA $(^\circ)$        & 104   & 103 $\pm \substack{5\\5}$\\
    $i$ $(^\circ)$       & 83       & $>61^b$          & $i$ $(^\circ)$         & 84    & $>51^b$ \\
    Ln prob              & -3400145.9 &                                &                      &       -3400147.0\\
    \hline
    \end{tabular}
    \vspace{.4cm}
    
     NOTE -- $^a$ A compact structure is not detected in the models with an extended structure, so the best-fit brightness values are not meaningful. The upper limits of 0.08 and 0.13 mJy represent the 99.7th percentile of the posterior distribution.
     
     $^b$ The lower limits on inclination of 61$^\circ$ and $51^\circ$ represent the 0.3rd percentile of the posterior distribution.
    \label{tab:mcmc2}
\end{table}

\section{Discussion} \label{discussion}
\subsection{Disk Structure Constraints}
\label{sec:disk_structure}
Our best-fit values for the disk's radius agree with previous determinations at optical wavelengths. \cite{lagrange16} and \cite{crotts21} find the radius of peak flux to be $\sim 64$ au, consistent with our determination of $\sim 60$ au. Our model indicates a radially broad ring, with a FWHM of $110 \pm \substack{50\\40}$ au. This contrasts with the vertical profile determined in \cite{crotts21}, since radially broad rings generally result in an increase in vertical FWHM with distance. However, it is consistent with the findings of their forward modeling, which indicated a broad ring. The high inclination of the disk and the limited SNR of the observations make it difficult to constrain the inner radius, which explains why the FWHM is poorly constrained.

In the larger-scale scattered light images, the HD~106906 debris disk shows an extreme brightness asymmetry, with the western side of the disk extending to nearly 6\farcs0 and the eastern side reaching only $\sim 4\farcs0$ \citep{kalas15}, while GPI images of the inner disk find the eastern extension to be $10-30\%$ brighter than the western extension. We do not see evidence of these features in our ALMA image, implying that the asymmetries are primarily present in the small, micron-sized grains that dominate the scattered light images. 

Somewhat surprisingly, the MCMC does not yield any evidence of a significant stellar offset relative to the center of the disk. The stellar offset we obtained from the image domain of $18 \pm 6$\,au along the disk midplane is similar to (and in the same direction as) the stellar offset of $\sim 16.5$\,au obtained from the near infrared observations via spine fitting \citep{crotts21}. Our MCMC likely prefers a small stellocentric offset due to the lack of asymmetry in the extended flux of the disk. Although the best fit narrow ring model, which is less sensitive to extended flux, has a slight stellar offset, that offset is not statistically significant. 

A truly eccentric disk is not the only explanation for a stellar offset, since disk material orbits around the center of mass of the system, which is not necessarily the star. The presence of a high-mass, high-separation planet allows the path of HD~106906AB to have a relatively large semi-major axis, up to $\sim 10$\,au for the 99.7th percentile of planet semi-major axis from \cite{nguyen21}. The highest values for the binary's semi-major axis correspond to the high-$a$, low-$e$ region of the planetary orbit's parameter space.  This situation would result in a stellar offset comparable to that suggested by the GPI imaging and image-domain ALMA analysis, but would not result in a brightness asymmetry at long wavelengths. It may result in some brightness asymmetry at short wavelengths, where brightness is dominated by scattering and scales steeply with proximity to the central star. Since HD~106906b is to the northwest of the binary, the stellocentric offset of the system's center of mass would also be to the northwest, which agrees with the observations. In such an arrangement, disk particles would not be stable in the region of the binary's reflex motion. Although the disk's inner edge is unresolved, some cavity around the disk's center is consistent with the lack of centrally peaked morphology in the ALMA image.

There are several limitations within our models that could impact our results. The first is that none of our MCMC models were inherently eccentric or asymmetric, other than fitting the peaks independently as point sources. While we are able to offset the disk's center from the star, we assume the disk to be circular with a symmetric dust density profile, which would not be the case for a truly eccentric disk. Given the lack of statistically significant residuals in our MCMC runs as well as the apparent brightness symmetry of the two peaks, it is unlikely that using a model that is eccentric with an asymmetric dust density profile would yield any additional evidence of eccentricity. Another limitation is that we are not able to fit the power laws of the inner and outer disk edges independently, and instead only model the disk as a Gaussian ring or a similarity solution with the outer power law fixed relative to the inner power law. This means that the since the inner radius is unresolved, we are unable to place a strong constraint on the location of the disk's outer edge.

\subsection{Dynamics of the Central Binary and External Perturber}
\label{sec:four-body}
Using the  constraints on the disk's morphology from the ALMA and GPI observations, along with the astrometric constraints on the orbit of HD~106906b, we can investigate the dynamical history of the system. While previous analysis has focused on the potential perturbing effect of the planet, the dynamical effect of the central binary remains largely unexplored. 

Any planetesimal in the disk experiences a gravitational perturbation both from the two stars in the binary and from the planet. Hence, determining the dynamical evolution of a planetesimal is essentially a four-body problem. Recent work simulating disk (self-) gravity has shown that it can likely have a significant effect on disk morphology, if the disk is sufficiently massive \citep{sefilian19, sefilian21}. However, given the relative masses and separations involved, the stellar binary and the HD~106906b companion are likely to dominate the gravitational potential at the location of the disk in this system. Although it is difficult to constrain the underlying planetesimal mass for debris disks, the observed flux indicates that the total mass of the disk around HD~106906 is likely on the order of 10\,M$_\mathrm{Earth}$ \citep{krivov21}. This is very low compared to the range of disk masses which yielded substantial dynamical effects in \cite{sefilian21}, $10^{-3}\leq M_d/m_p \leq1$.
We can begin to disentangle the dynamical effect of the planet from the dynamical effect of the binary by comparing their precession timescales as a function of location in the disk.

We use expressions for the procession frequency caused by the binary ($A_B$) and the planet ($A_p$) derived in \cite{rodet17}:
$$A_B = \frac{3}{16}n\left(\frac{a_B}{a}\right)^2\left(\frac{3}{2}e_B^2 + 1\right)$$
and
$$A_p = \frac{3}{4}\frac{m_p}{m_B}n\left(\frac{a}{a_p}\right)^3(1-e_p^2)^{-\frac{3}{2}}$$
where $n$ is the period of the disk particle, $a$ is the semi-major axis of the disk particle, $a_B$ and $a_p$ are the semi-major axis of the binary, planet, $e_B$ and $e_p$ are the eccentricity of the binary and planet, $m_p$ is the mass of the planet, and $m_B$ is the combined mass of the binary. These expressions were derived assuming that the binary mass parameter is very close to 1/2, treating disk particles as massless test particle, and ignoring the third order effect of the binary, since $a_B << a$.

\begin{figure}
    \centering
    \includegraphics[scale=0.7]{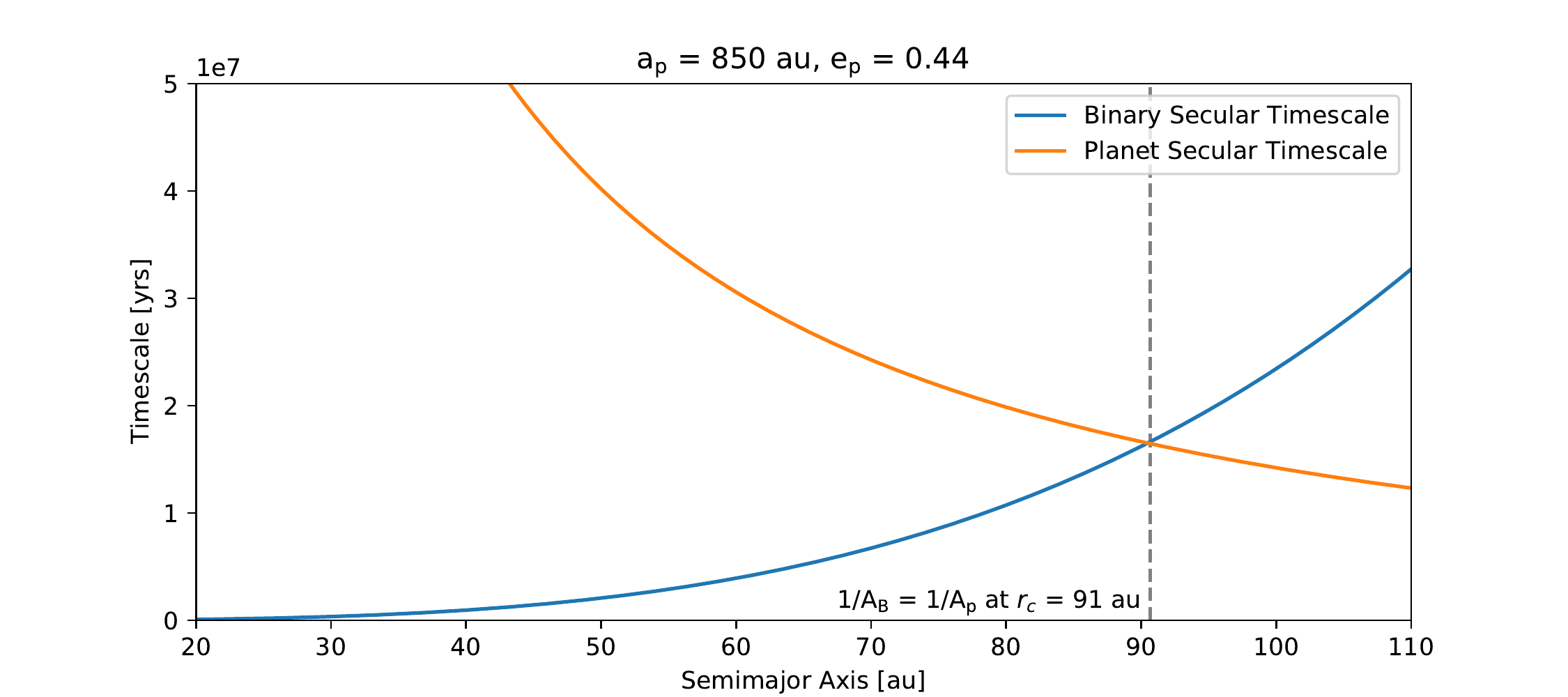}
    \caption{Timescale of secular interaction with the binary ($\frac{1}{A_\text{p}}$) in blue and the planet ($\frac{1}{A_\text{B}}$) in orange over the region of the disk for $a_\text{p} = 850$\,au and $e_\text{p} = 0.44$. The dashed gray line represents the radius, $r_c$, where the two timescales are equal. Interior to that radius,  interaction with the binary will dominate the disk's morphology, while exterior to that radius interaction with the planet will dominate.}
    \label{fig:rodet1}
\end{figure}

Assuming $a_B = 0.367$\,au, $e_B = 0.669$, and $m_B = 2.71$ M$_\odot$ \citep{derosa19}, we can determine the secular timescale of interaction with the binary as a function of $a$. By comparing this timescale with the timescale for interaction with the planet, we can get a rough estimate of which disk regions will be shaped by the binary or the planet. Figure\,\ref{fig:rodet1} shows the timescale of secular interaction with the binary and the timescale of secular interaction with the planet for the median planet semimajor axis and eccentricity found by \cite{nguyen21}. The dashed gray line shows the radius where the two secular timescales are equal. Interior to that radius, the disk's morphology is largely determined by interaction with the binary. Exterior to that radius, the disk's morphology is largely determined by interaction with the planet. Since the disk's brightness peaks near a radius of $\sim 60$\,au and most of the brightness is interior to $\sim 100$\,au, we can assume that the morphology of the disk would mostly be controlled by the binary for this configuration of planet parameters.

\begin{figure}
    \centering
    \includegraphics[scale=0.5]{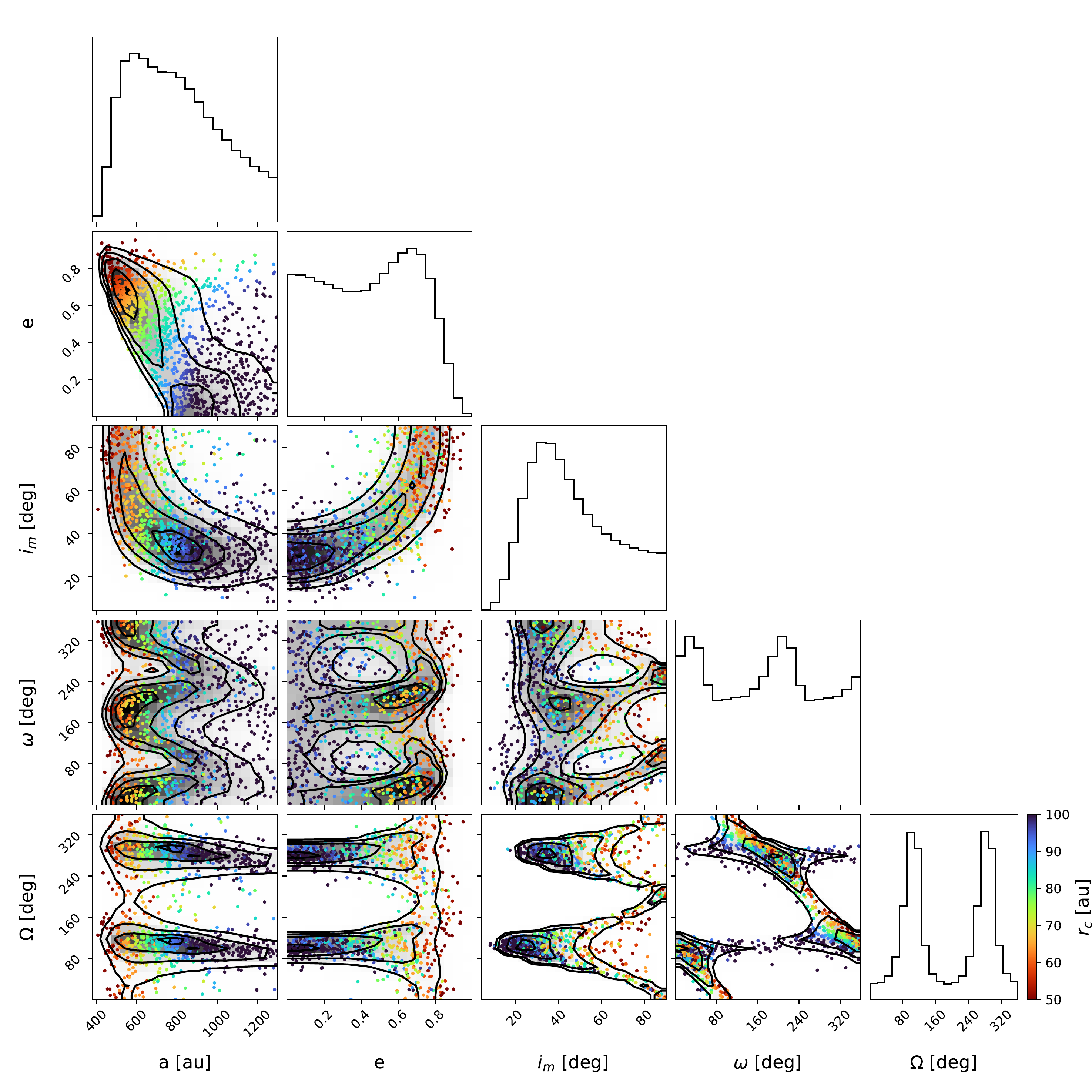}
    \caption{Corner plot based on posterior distribution from \citet{nguyen21}, showing five orbital parameters for HD~106906b. In order of appearance, the parameters included are (1) semi-major axis, $a$; (2) eccentricity, $e$; (3) mutual inclination $i$; (4) argument of periastron, $\omega$; and (5) longitude of the ascending node, $\Omega$. The gray scale probability distributions show 2D covariances and 1D marginalized posterior distributions derived from astrometric modeling. The colored dots are a random sample of the posterior distribution, with color representing the semi-major axis where the secular timescale of the planet is equal to the secular timescale of the binary ($r_c$).  The regions of parameter space where the binary stabilizes the disk, leading to morphology consistent with the ALMA data, are characterized by large $a$, low $e$, and low $i_m$. }
    \label{fig:rc_corner}
\end{figure}
 To investigate the relative influence of the central binary and the planet on the disk as a function of the orbital properties of HD~106906b, we calculate the transition radius ($r_c$) where $\frac{1}{A_B} = \frac{1}{A_p}$ across the planet's orbital parameter space. Figure \ref{fig:rc_corner} shows $r_c$ for 1000 points selected from the posterior distribution from \cite{nguyen21}. A critical radius greater than $\sim$ 100 au indicates that the disk's morphology is dominated by interaction with the binary, while a critical radius less than $\sim$ 60 au indicates that the disk's morphology is dominated by interaction with HD~106906b. For the majority of planet orbits allowed by the astrometric constraints, the binary dominates in the inner region of the disk, while the planet dominates in the outer region, with $r_c$ falling interior to the disk's outer edge of $\sim$100\,au.
 
 In the next sections, we examine the dynamical effect of the planet alone on the disk (Section~\ref{sec:planet}), and then carry out some N-body simulations -- both with and without the central binary -- to verify the intuition provided by our analytical investigation (Section~\ref{sec:n-body}).  
 
\subsection{Dynamical Effect of the Planet}
\label{sec:planet}
First, we analyze the dynamical effect of the planet by neglecting the binary nature of HD~106906 and considering dynamical constraints from the literature. We consider two conditions for stability in triple systems, the results from \cite{eggleton95} and \cite{mardling01}. Although there are many particles in the disk, we consider each particle to be in a triple system consisting of a central body (representing the binary), the planet, and the test particle. 
\cite{eggleton95} derived an analytical stability boundary, where a system is considered stable if the initial hierarachy of semi-major axes is preserved and no bodies are ejected for 100 orbits.
\cite{mardling01} derived an expression for stability by analogy to chaotic energy exchange in the binary-tides problem. 
We adopt the versions of the equations rewritten in \cite{he18}, which uses
$$r_{ap}\equiv\frac{a_{\text{p}}(1-e_{\text{p}})}{a(1+e)}$$
as the orbital separation parameter. In our case, we consider $a$ to be the radius of peak disk brightness, which we take as the median $R_0$ of the Gaussian ring fit, $\sim$60 au, and $e$ is the initial eccentricity of the disk particles, which we assume to be negligible.
The system is stable when $r_{ap}>Y$, with $Y_{EK}$ and $Y_{MA}$ to denote the results from \cite{eggleton95} and \cite{mardling01} respectively as
$$Y_{EK} \equiv 1 + \frac{3.7}{q^{1/3}}- \frac{2.2}{1+q^{1/3}}$$
$$Y_{MA} \equiv 2.8 \frac{1}{1+e}\left[\left(1+\frac{1}{q}\right)\frac{1+e_{p}}{(1-e_{p})^{1/2}}\right]^{2/5}(1-0.3i_m/180^{\circ}).$$
where $q = \frac{m_B}{m_p}$, the mass ratio of the binary to the planet, and $i_m$ is the mutual inclination of the planet and the disk midplane. The expression for $Y_{EK}$ also includes an additional term relating to the mass ratio of the two outer bodies, which we ignore since one of them is the test particle.

\begin{figure}
    \centering
    \includegraphics[scale=0.6]{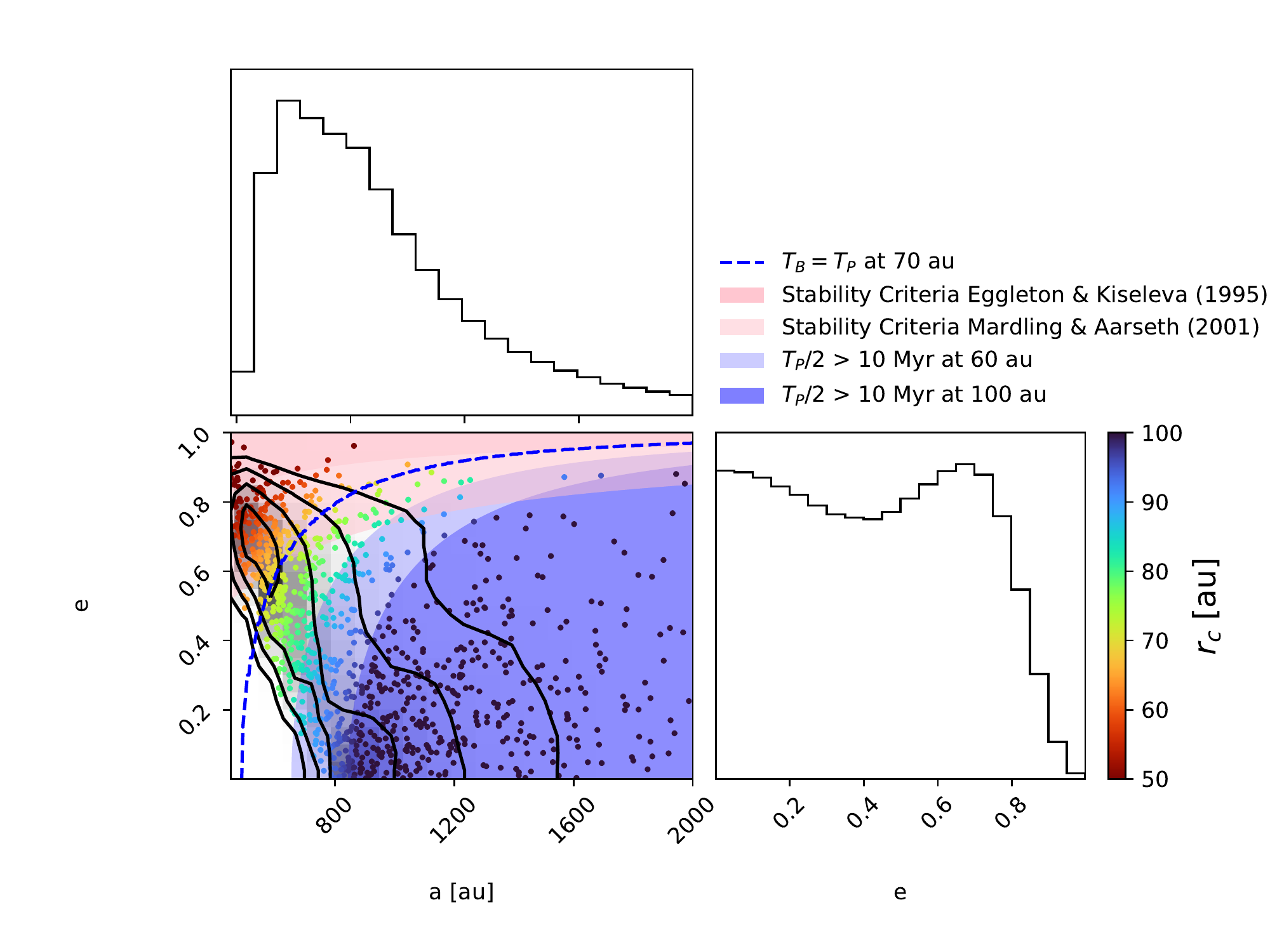}
    \caption{Posterior distribution of planet semi-major axis and eccentricity from \cite{nguyen21}. The gray scale histogram and scatter points are as in Figure \ref{fig:rc_corner}.  The pink shading show the stability criteria $Y_{EK}$ and $Y_{MA}$, and the blue shading shows the region of parameter space where the secular timescale of the planet is too long for dynamical interactions to take effect within the lifetime of the disk. The dashed blue line shows the location where the secular timescale of interaction with the planet is equal to the secular timescale of interaction with the binary at 70\,au. Above this line, the morphology of the disk at the radius of peak brightness is primarily controlled by interaction with the planet. The region of parameter space where the disk is primarily shaped by the planet is also the region of parameter space where the disk is likely unstable.}
    \label{fig:planetstability}
\end{figure}

The pink shading in Figure \ref{fig:planetstability} shows the range of planetary semi-major axes $a$ and eccentricities $e$ where the disk would be unstable in the absence of the binary. Within the pink shading, we can assume that the disk will be highly disrupted exterior to $r_c$. A majority of the planetesimals in that region will either be ejected, or have a large induced eccentricity and mutual inclination. Since the ALMA emission indicates a symmetric disk and the scattered light observations indicate a vertically flat disk, a dynamically unstable disk is unlikely to be consistent with the observed morphology. In Section \ref{sec:n-body}, we confirm this conclusion with $n$-body simulations of the system.

We also consider the secular timescales of interactions with the planet compared to the age of the system. In order for a region of the disk to be maximally disrupted within some time $t$, that part of the disk must have experienced at least one half-cycle of secular perturbation, so the secular timescale at that location in the disk must be at most $\frac{1}{2}t_p \lesssim t$, where $t_p = \frac{1}{A_p}$. In Figure \ref{fig:planetstability}, the blue shading shows the region of the parameter space where perturbations from the planet would not reach the outer edge of the disk (100\,au) or the region of peak brightness (60\,au) within 10\,Myr. In this region of parameter space, we would expect to see very little induced eccentricity or mutual inclination in the disk from interaction with HD~106906b, regardless of the behavior of the binary.  This lack of perturbation is consistent with the small scale height and symmetric morphology of the GPI and ALMA data, indicating that models within the blue-shaded region of Fig.~\ref{fig:planetstability} are consistent with the observational constraints on the planetesimal ring. These constraints may loosen depending on the how long HD~106906b has been on its current orbit. Although the age of the system is 13$\pm2$\,Myr \citep{pecaut16}, it is possible that the planet formed relatively recently, or that it was on a different orbit prior to a scattering event caused by interaction with the binary, a stellar flyby \citep{rodet17}, or a free-floating planet (Moore et al. in prep). If the planet only arrived on its current orbit within the past few million years, a much more eccentric, closer-in orbit would be allowed.

The dynamics of the system, and particularly the timescale of its evolution, depend strongly on the mass of the planet, $11.9\pm\substack{1.7\\0.8}$\,M$_\mathrm{Jup}$ \citep{daemgen17}. Within the 1-$\sigma$ margin of the planet's mass, the timescale of interaction with the planet can vary by a few million years. If the planet were more massive, it would be more disruptive on the same orbit. However, even considering alternate evolutionary models, the mass of HD~106906 is at most $\sim14_\mathrm{Jup}$, the effect of which is small compared to both the uncertainty on the planet's orbit and the disk's morphology.

Two studies have previously simulated interactions between HD~106906b and its debris disk.  \cite{nesvold17} simulated the HD~106906 system with a low mutual inclination ($i \leq 30^\circ$) and high eccentricity ($e = 0.7$) orbit for HD~106906b. Their results suggest that a mutual inclination of $i = 30^\circ$ was sufficient to induce significant vertical extension, and hence prefer a planet with mutual inclination less than $\sim 10^\circ$, due to the lack of vertical extension present in scattered light images \citep{kalas15,lagrange16}. However, the high-eccentricity/low-mutual-inclination parameter space is largely excluded by the astrometric constraints on the planet's orbit from \cite{nguyen21}, which suggest that an eccentricity as large as $\sim0.7$ would require a high mutual inclination, $\sim60^\circ$ to be consistent with the astrometric data. Additional constraints on vertical extent were obtained in \cite{crotts21}, who were able to measure the scale height of the disk for the first time and obtain an intrinsic vertical FWHM of $\sim0\farcs15$ ($\sim15.6$\,au). Also, the disk model resulting from the \cite{nesvold17} simulation would result in brightness asymmetry in the inner disk that is not present in the ALMA observations. Hence, in agreement with the stability criteria, the results of \cite{nesvold17} suggest that long-term high-eccentricity, high-inclination orbits are inconsistent with the observational constraints.

\cite{rodet17} simulated an ejection scenario for  HD~106906b, and found that the passage of the planet through the disk induces an asymmetry in the outer disk ($\sim500$\,au) that is qualitatively consistent with observations at optical wavelengths. As the system continues to evolve after the scattering event, secular interaction with the binary will obscure artifacts of its effects, first in the inner disk and then propagating outward. Hence, an ejection scenario could cause asymmetry in the outer disk without an asymmetric inner disk, which is consistent with the lack of asymmetry in the ALMA observations, but not with the brightness asymmetry determined in \cite{crotts21}. This type of scattering event would require a stabilizing interaction to put the companion back on a stable orbit, possibly in the form of a stellar flyby. \cite{derosa19} investigate possible flyby candidates, and identify HIP 59716 and HIP 59721 as the most likely candidates for a close encounter. \cite{rodet19} simulate possible flyby interactions, and find that the two candidates could have had some dynamical interaction with HD~106906b or the disk, but a significant impact on their orbit is unlikely. However, this does not rule out the possibility of a stabilizing interaction with some other body that has not been identified or detected, e.g., a free-floating planet as proposed by Moore et al.~(submitted). It is also possible that the planet's orbit was circularized via interaction with disk material, if the outer disk was much more massive at the time of the planet's ejection than it is today. The low probability of a stabilizing interaction, along with the properties of the sample of PMCs described in \citet{bryan2016} makes a scattering event unlikely, but it cannot yet be ruled out conclusively for this particular system. 

\subsection{Dynamical Simulations}
\label{sec:n-body}
To determine whether the timescales and stability criteria discussed above accurately characterize the particular  architecture of the HD~106906 system, we ran $n$-body simulations using the software package {\tt REBOUND} \citep{rein12} with hybrid integrator {\tt MERCURIUS}, which switches from a fixed to variable timestep to evaluate close approaches involving a body with mass.

Particles are initially placed on random orbits, with a uniform distribution in semi-major axis, $a$, between 10 and 500 au, in order to test the ability of  the star and the planet to sculpt the inner and outer edge of the disk, respectively. Their initial eccentricities are uniformly distributed between $0-0.02$, and all particle orbits are initialized in the disk midplane. We use $10^4$ particles, sufficient to sample the span in semi-major axes and recover smooth images of the disk density distribution. 
The planet's orbital parameters were selected randomly from the posteriors of the astrometric constraints obtained in \cite{nguyen21}. The planet is given a mass of 11 $M_\text{Jup}$ and a bulk density of 1.64 g cm$^{-3}$ (Neptune's density). Varying the mass of the planet would effect the rate at which the orbit of disk particles evolve, as well as the magnitude of perturbations. However, the uncertainty on the planet's mass is small compared to the range of planet orbital parameters we explore. The radius of the planet has no effect on the morphology of the disk, since the planet does not pass through the disk, and the region where the planet's radius would become relevant is unresolved. We integrate the evolution of the system up to 10\,Myr, similar to the age of the system.
We run two sets of simulations, first using a single particle to represent the binary, and secondly including both stars in the simulation. In the single-mass case, the fixed timestep is chosen to be 15\% of the period of the innermost disk particle, since all particle periods are much shorter than the planet's period. In the binary case, the fixed timestep is chosen to be 15\% of the binary's orbit at its closes approach, i.e., 15\% of the binary's period if its semi-major axis was its pericenter distance. Both because of the addition of another body with mass and the short period of the binary, simulating dynamical interaction with the binary proved to be extremely computationally intensive, so we only ran simulations with the binary for a couple of carefully selected cases in order to verify the intuition provided by the dynamical analysis. Although the dynamical effect of a binary can sometimes be simulated using a J2 quadrupole potential, doing so would poorly represent the HD~106906AB system due to its high eccentricity \citep[$e=0.669\pm0.002$;][]{derosa19}.

\subsubsection{Disruption by the Planet}
\begin{figure}
    \centering
    \includegraphics[scale=0.55,trim={2cm 2cm 2cm 0cm},clip]{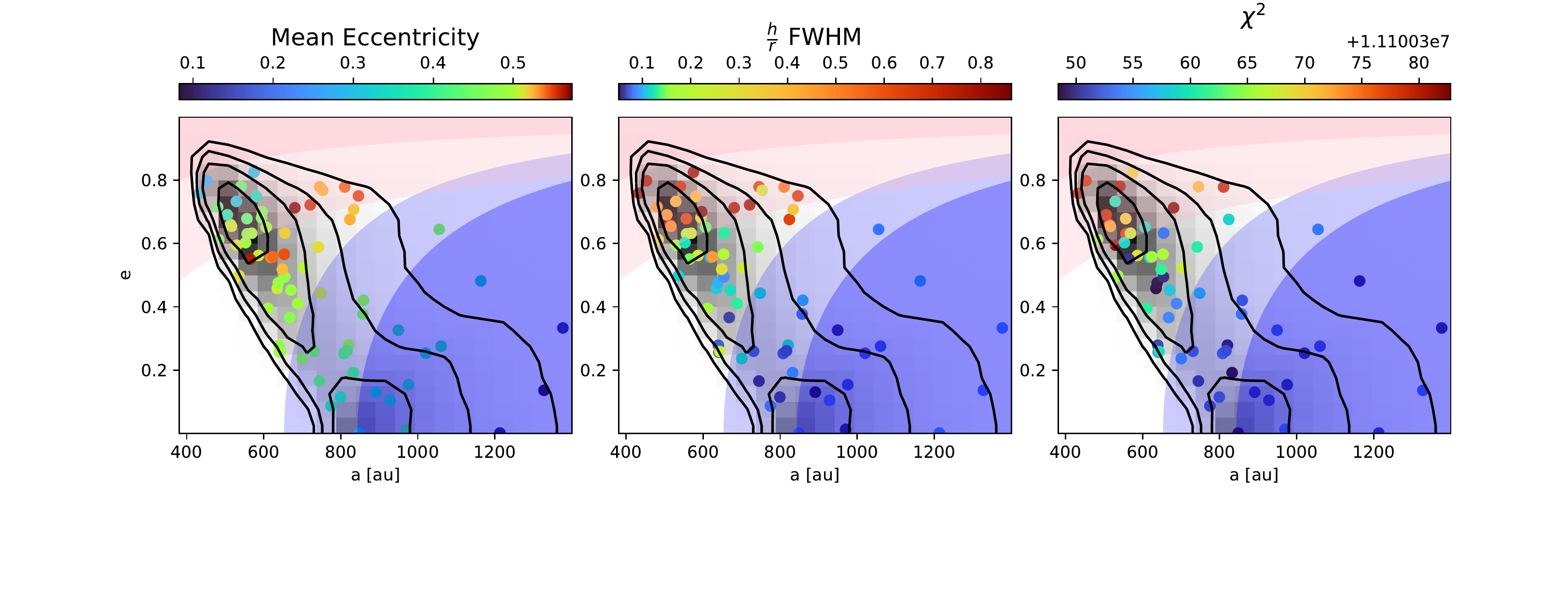}
    \caption{Mean particle eccentricity (left), $h/r$ FWHM (center), and $\chi^2$ (right), for simulated disks with a variety of planet orbital parameters after 10\,Myr of integration are shown by the colored scatter points. The gray scale histogram shows the posterior distribution of planet semi-major axis and eccentricity from \citep{nguyen21}. The colored shading is as in Figure \ref{fig:rc_corner}.}
    \label{fig:planetsim}
\end{figure}
When simulating the system with a single object representing the binary, the mass of the central particle is the combined mass of the binary, 2.71\,$M_\odot$. We can use the results of these simulations to evaluate the dynamical criteria discussed in Section \ref{sec:planet}. 

First, we compared the simulations to the observations by investigating the evolution of disk particle orbital parameters. This allows us to evaluate each set of planetary orbital parameters while ignoring many of our assumptions for initial disk parameters. The mean eccentricity of disk particles for each simulation is shown in the leftmost panel of Figure \ref{fig:planetsim}. In the region of parameter space deemed unstable by dynamical criteria, a large fraction of the initial disk particles are ejected from the system, which explains the more moderate eccentricities of the remaining disk particles in that region of parameter space. The center panel of Figure \ref{fig:planetsim} shows the vertical FWHM of the disk material after integration. In the region of parameter space where more than 1/2 of a secular timescale has passed, the FWHM is too large to be consistent with the scattered light constraints.

We also produced synthetic ALMA observations of the systems to compare directly with the data. In order to translate the outcome of the simulations to density distributions used to produce synthetic images, we populate the orbit of each particle with 200 points distributed uniformly in mean anomaly. For each simulation, we then obtain a best fit initial surface density by weighting the mass of each particle based on its initial semi-major axis. We assume an initial surface density proportional to $r^\gamma$ between some inner and outer radius, $r_{\text{in}}$ and $r_{\text{out}}$.  For each simulation, we then adjust the initial surface density by weighting the mass of each particle based on its initial semi-major axis and fit for $\gamma$, $r_{\text{in}}$, and $r_{\text{out}}$ to best match the data. We use the {\tt galario} function {\tt sampleImage} to convert our density distributions into synthetic ALMA visibilities. We then compare these visibilities to the observed visibilities by calculating a $\chi^2$ value for each simulation, which are shown in the leftmost panel of Figure \ref{fig:planetsim}. None of the perturbed disks yielded a $\chi^2$ value significantly lower than that of the initial axisymmetric distribution; in general, the more perturbed a disk was, the higher its $\chi^2$ value. The distribution of $\chi^2$ values generally prefers planets on less eccentric orbits with higher semi-major axes, which have more distant periastra and hence longer timescales for interaction with disk material. That is, the $\chi^2$ analysis prefers an unperturbed disk that remains axisymmetric for the duration of integration.

\subsubsection{Stabilizing Effect of the Binary}

\begin{figure}
    \centering
    \includegraphics[scale=0.6]{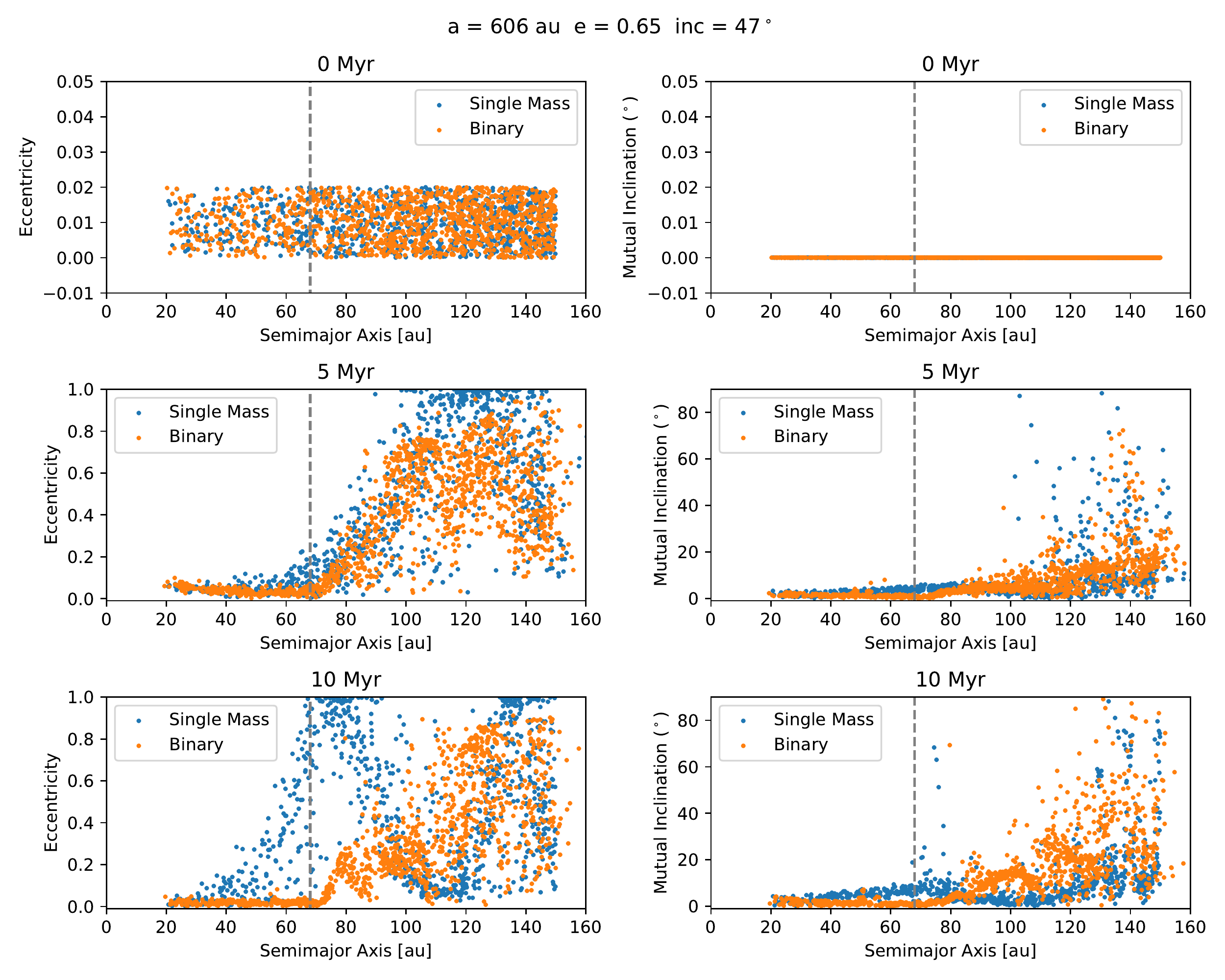}
    \caption{Evolution of disk particles in an example simulation with a single mass representing HD~106906 (in blue) and with the binary (in orange). The left column shows the eccentricity of the particles, where the right column shows the inclination of the particles relative to the disk midplane. The top row is the initial state of the disk, the second row is after 5\,Myr of integration, and the bottom row is after 10\,Myr of integration. The dashed gray line shows the critical radius, $r_c$, calculated by comparing the secular timescales of interaction with the binary and the planet. The disk particles are stabilized by the binary interior to that radius, so they have very low eccentricities and mutual inclinations after integration.}
    \label{fig:evol}
\end{figure}

\begin{figure}
    \centering
    \includegraphics[scale=0.54,trim={2cm 0 4cm 0},clip]{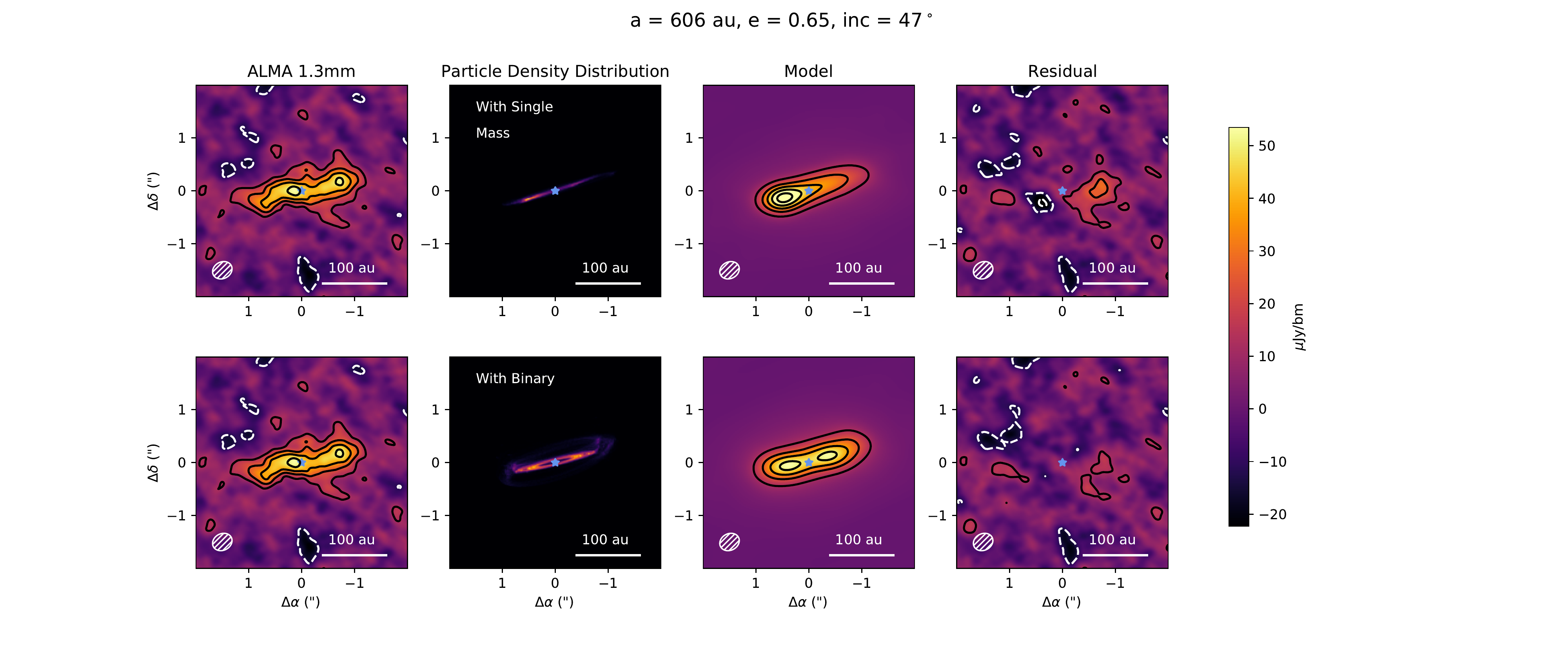}
    \caption{Simulated disk morphology of an example simulation after 10\,Myr of integration. The top row corresponds to a simulation where  HD~106906 is represented by a single body, while the bottom row corresponds to a simulation where HD~106906 is represented by a binary. The left panel of each row shows the ALMA observations, the second panel shows the density distribution of simulated disk particles, the third panel shows a synthetic ALMA image of the simulated disk, and the final image shows residuals. For the 1st, 3rd, and 4th panels countour levels and symbols are as in Figure \ref{fig:1}.}
    \label{fig:nbody_dmr}
\end{figure}

To evaluate the dynamical effect of the binary, we implement it in our simulations with $e = 0.669$ and $a = 0.37$ \citep{derosa19}, and with masses of $m_A = 1.37$\,M$_\odot$ and $m_B = 1.34$\,M$_\odot$. The inclination of the binary's orbit is unknown, so we initially assume that it is coplanar with the disk midplane. We investigate a range of inclinations for the binary in Section \ref{sec:bin-inc}.

By comparing the simulations with and without the binary, we can see a striking difference in the region of the disk where the secular timescale of interaction with the binary is less than or similar to the secular timescale of interaction with the planet. 
In this region of the disk, particle orbits appear unperturbed throughout the integration of the system, in the presence of the binary. Their orbits remain essentially circular and close to the midplane of the disk. Figure \ref{fig:evol} shows the evolution of disk particles in a simulation with a single central point mass and with a central binary for an example simulation with a planet orbit of $a=606$\,au, $e=0.65$, and $i=47^\circ$. The dashed gray line shows the analytical $r_c$ for this set of planet parameters. After 10\,Myr, there is clearly significant induced eccentricity in the disk that is inconsistent with the symmetric ALMA brightness in the absence of the binary. However, the binary is able to stabilize disk particles out to $r_c$.

To determine what range of $r_c$ values was consistent with the ALMA observations, we also produced synthetic visibilities for the simulations including the binary, using the method described above. An example simulation is shown in Figure \ref{fig:nbody_dmr}, both for the single mass (top) and binary (bottom) case.  The orbital parameters of the planet ($a$=606\,au, $e$=0.65, $i$=47$^\circ$) were chosen to illustrate a case where the critical radius of transition from binary-dominated to planet-dominated particles falls at 68\,au, just outside the radius of peak brightness of the disk.  For this simulation, the orbital parameters produce a highly eccentric disk that is inconsistent with the ALMA data if we treat the central binary as a point mass (top row). However, when the stabilizing effect of the binary is taken into account, the simulation produces a disk that is in fact consistent with the ALMA observations (bottom row), as illustrated by the lack of statistically significant ($> 3 \sigma$) residuals. Hence, the region of parameter space where the radius of transition is greater than $\sim$70\,au, marked with a black line on Fig.~\ref{fig:planetstability}, seems to be consistent with the ALMA observations.

It is more difficult to determine whether this stabilizing effect will hold for the material that is visible in the scattered light. Although there will likely be some stabilizing effect for those particles as well, this may be offset by the effect of radiation pressure on smaller grains (Moore et al. in prep).

\subsubsection{The Binary Orbital Plane}\label{sec:bin-inc}
\begin{figure}
    \centering
    \includegraphics[scale=0.6]{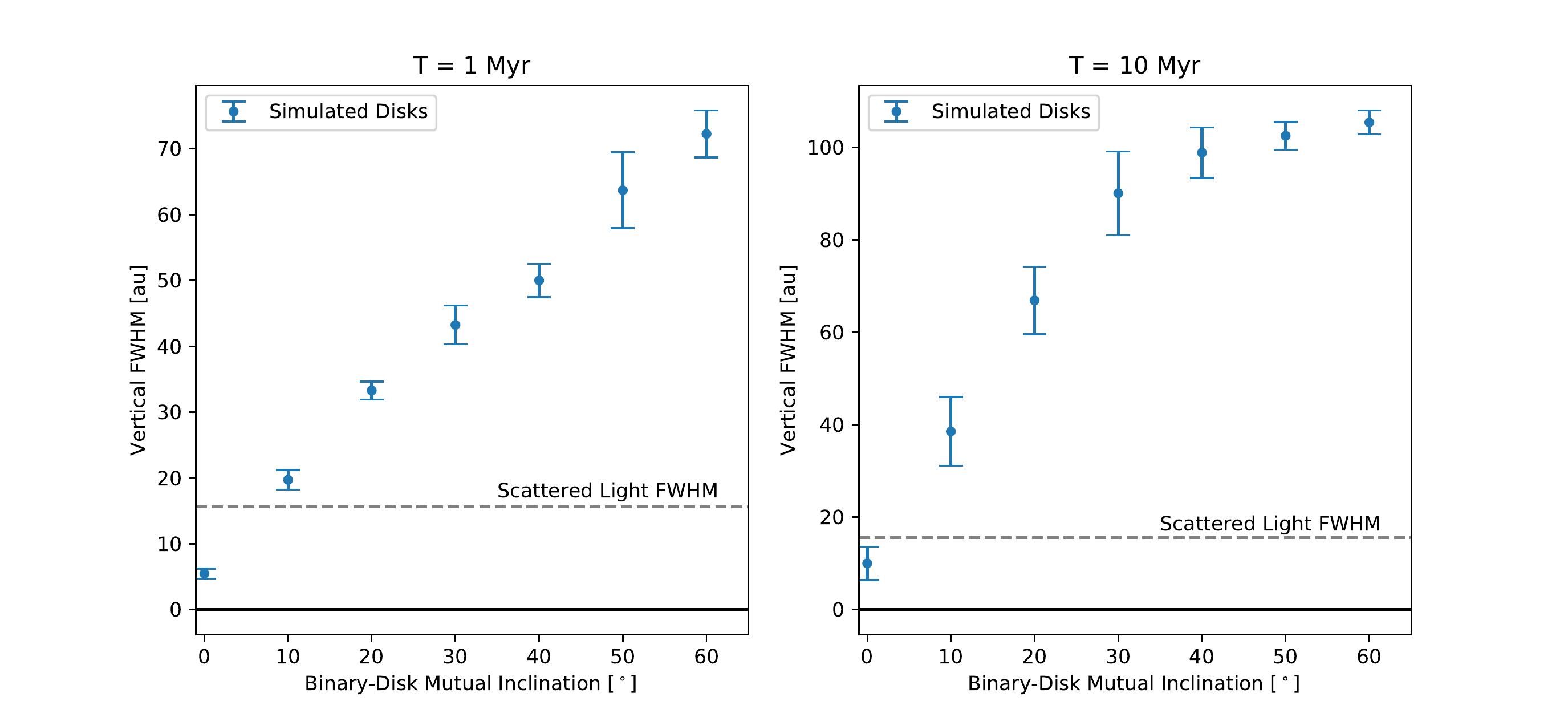}
    \caption{Vertical FWHM after 1\,Myr (left) and 10\,Myr (right) of integration for  a range of mutual inclinations between the binary orbital plane and the disk midplane. Each blue scatter point represents the mean FWHM for a set of 5 simulations with vary planet parameters, with error bars showing the standard deviation.}
    \label{fig:binary_inc}
\end{figure}

In the case where the binary is initially coplanar with the disk midplane, we do not expect that it would induce any scale height in the disk. However, there are several observed cases of circumbinary disks that are misaligned with respect with respect to the binary orbital plane, e.g. KH\,15D \citep{winn04}, IRS\,43 \citep{brinch16}, and GG\,Tau\,A \citep{aly18}. Previous studies have investigated the dynamics of accreting protoplanetary disks around initially misaligned eccentric binaries \citep[e.g.,][]{facchini13, martin17, zanazzi18, smallwood19, martin22} and found that viscous torque will generally either align a circumbinary disk with the binary's orbital plane or lead to a disk configuration perpendicular to the binary orbit. In addition, \cite{czekala19} found that the majority of circumbinary disks around short-period binaries are nearly coplanar, with mutual inclinations less than $\sim3^\circ$.

Although the case where the binary's orbital plane is inclined relative to the disk at the debris disk phase is likely non-physical, we can also place a constraint on the misalignment by simulating the system with a range of binary inclinations and comparing with the observed morphology. For a set of 5 different planet parameters drawn from across the posterior distribution from \cite{nguyen21}, we simulate the system with a variety of mutual inclinations between the binary and the disk for 10\,Myr. Figure \ref{fig:binary_inc} shows the maximum FWHM of the simulated disk at $\sim60$\,au after 1\,Myr and 10\,Myr of integration. By comparing with the intrinsic FWHM of 15.6\,au derived in \cite{crotts21}, we can rule out misalignment between the binary orbital plane and the disk midplane. In particular, we can rule out a case where the disk is coplanar with the orbit of HD~106906b. Although the binary has likely had the same separation, eccentricity, and inclination since the formation of the system, its dynamical effect on the disk  is impacted by the orbit of the planet, and particularly by the resulting reflex motion of the binary around the system's center of mass.
Hence, the simulations only accurately represent dynamical interaction since HD~106906b arrived on its current orbit, which may have happened relatively recently.
This could relax our constraint on the binary's inclination.

\subsection{Formation of HD~106906b}

The ALMA observations, in concert with the \textit{HST} astrometry and GPI scattered light, effectively rule out a long-term high-$e$, low-$a$ orbit for HD~106906b, and instead favor a large-$a$, low-$e$, nearly circular orbit. We also place low upper limits on the mass of a circumplanetary disk around HD~106906b.  These results should be considered within the context of the proposed formation scenarios for HD~106906b; in particular, the observational constraints so far seem at least plausibly consistent with either formation in situ via gravitational instability, or with a scattering event that ejected the planet from the disk.  

In the case of in situ formation, the large, low-$e$ orbit favored by the observational constraints would have persisted from the time of formation of the planet.  The misalignment between the planet's spin axis and its orbital plane \citep{bryan21} would result from the initial fragmentation of the disk and random variations in the angular momentum of the progenitor cloud.  The lack of emission from the circumplanetary disk is perhaps surprising in this scenario, since we might expect a robust circumplanetary disk to extend to a substantial fraction of the planet's relatively large Hill radius at such a large separation from its host star.  However, it would be consistent with the overall low detection rates of circumplanetary material around PMCs despite the presence of red colors similar to those observed for HD~106906b -- a discrepancy which is still not well understood for this comparable sample of objects \citep{wu2017,wu2020}.  The tracers of a circumplanetary disk are particularly ambiguous in the case of HD~106906b, including the lack of spectroscopic accretion signatures \citep{daemgen17}, alongside the tentative evidence for marginally resolved structure in the \textit{HST} observations \citep{kalas15}.  

The alternative scenario is that HD~106906b formed within the circumbinary disk and was scattered onto its current orbit, either via interactions with the binary and a stellar flyby or due to a close encounter with a free-floating planet \citep[][Moore et al. in prep]{derosa19}.  In this case, the lack of circumplanetary material would be entirely consistent, both since objects closer to their host stars have smaller Hill radii (and circumplanetary disks are generally accrued early in a planet's life), and since some of the disk material may have been stripped during the ejection event.Ejection could possibly induce misalignment between the spin axis of the planet and the disk plane, as observed by \citet{bryan21}, although it is unclear whether it could produce the particular angular momentum architecture of this system.  The large-$a$, low-$e$ orbits for the planet preferred by the ALMA data are less likely in this scenario, although since the timing of the scattering event is unknown, the constraints on the planet's semimajor axis and eccentricity could be more relaxed if the event were relatively recent (e.g., within the past couple Myr).  A short-timescale scenario is most favorable for keeping the vertical height of the scattered light disk low, as observed in the GPI data \citep{crotts21}.  The large-scale asymmetries observed in scattered light may also favor this scenario, since they are easiest to induce with a higher-eccentricity orbit. However, this scenario remains unlikely due to the low probability of a stellar flyby or passage of a free-floating planet with sufficiently low impact parameter to induce the requisite changes in the planet's orbit.

\section{Summary and Conclusions} 
\label{summary}

As the only known system (to date) containing a debris disk along with a directly imaged external companion, HD~106906 presents a unique opportunity to study dynamical interactions between a planet and a disk. We present ALMA observations at a wavelength of 1.3 mm that spatially resolve the structure of the disk for the first time at millimeter wavelengths. The ALMA image shows one bright peak on either side of the binary, consistent with an axisymmetric flux distribution and a marginal positional offset between the disk center and the position of the binary. 

We fit models to the visibilities to robustly determine the geometry of the system. The best-fit models reveal an extended distribution of dust out to a distance of 100$\pm20$ au, with no statistically significant evidence of disk asymmetry or stellocentric offset. There is no evidence from the ALMA data for the dramatic east-west asymmetry seen in scattered light images. 

We perform a dynamical analysis, supported by N-body simulations, of long term disk-planet interactions, which indicate that high-$e$, low-$a$, or high-$i_m$ orbits for HD~106906b produce a highly disrupted disk on timescales comparable to or less than the age of the system -- a scenario that is inconsistent with the ALMA observations and with limits on scale height from the GPI data. Only the low-$e$, large-$a$ region of parameter space, where there is little to no secular interaction within the relevant timescales, reproduces the observed disk morphology.  The constraints are relaxed somewhat by the stabilizing influence of the central binary, although the highest-eccentricity region of parameter space from the astrometric constraints (where the radius of influence of the planet extends within about 70\,au) are still strongly inconsistent with the ALMA data on 10\,Myr timescales.  The reflex motion due to the large separation of the planet is sufficient to explain any offset of the midpoint between the peaks from the location of the binary, even with a circular disk. An eccentricity larger than 0.6 within a distance of 100\,au from the central star is ruled out by the brightness ratio between apo- and pericenter observed in the ALMA data, and low eccentricities are favored. 

While the origin of the scattered light asymmetry is still unclear, short term dynamical interactions with the planet on its current orbit seem a likely cause. While the lack of dust emission at the location of the companion is suggestive of a disrupted circumplanetary disk following ejection, it is also generally consistent with low upper limits on PMCs despite other lines of evidence for circumplanetary material.  At this point, the preponderance of the evidence seems to favor an in situ formation scenario, although a scattering scenario cannot be definitively ruled out for this system.

\begin{acknowledgments}
The authors gratefully acknowledge Meiji Nguyen for providing the posteriors for the {\it HST} astrometry.  AJF acknowledges support from the Wesleyan College of Integrative Sciences and NASA CT Space Grant Consortium.  AMH is supported by a Cottrell Scholar Award from the Research Corporation for Science Advancement.    

This paper makes use of the following ALMA data: ADS/JAO.ALMA\#2017.1.00979.S. ALMA is a partnership of ESO (representing its member states), NSF (USA) and NINS (Japan), together with NRC (Canada), MOST and ASIAA (Taiwan), and KASI (Republic of Korea), in cooperation with the Republic of Chile. The Joint ALMA Observatory is operated by ESO, AUI/NRAO and NAOJ.  This research made use of {\tt Astropy}, a community-developed core Python package for Astronomy \citep{astropy}, {\tt matplotlib} \citep{matplotlib}, {\tt numpy} \citep{numpy}, and {\tt pandas} \citep{pandas}.  This research has made use of NASA's Astrophysics Data System.  This research has made use of the AstroBetter blog and wiki.  This research has made use of the SIMBAD database, operated at CDS, Strasbourg, France.
\end{acknowledgments}

\bibliography{references}{}
\bibliographystyle{aasjournal}

\end{document}